\documentclass[aps,twocolumn,a4paper,showkeys]{revtex4-1}
\usepackage{amsmath,amssymb,bm,amsthm}
\usepackage{subfigure}
\usepackage{graphicx}
\usepackage{dcolumn}
\usepackage{bm}
\usepackage[mathlines]{lineno}
\usepackage{booktabs}
\usepackage{color}
\newcounter{one}
\usepackage{url}

\usepackage{textcase}

\usepackage{colortbl}
\usepackage{tabularx}
\usepackage{verbatim}
\usepackage{multirow}

\usepackage[T1]{fontenc} 
\usepackage{lmodern}
\usepackage{bbm}
\usepackage[utf8]{inputenc}
\usepackage{amsfonts}
\usepackage{array}

\usepackage{bbm}
\usepackage{enumitem}
\usepackage{umoline}
\usepackage[usenames,svgnames]{xcolor}
\usepackage{natbib}
\usepackage[hyperindex,breaklinks]{hyperref}
\hypersetup{
     colorlinks=true,       		
     linkcolor=Navy,          	
     citecolor=Navy,            
     filecolor=Navy,      		
     urlcolor=Navy,           	
    runcolor=cyan,
 }
\usepackage{graphicx}
\usepackage{amsfonts}
\usepackage{amssymb}
\usepackage{amsmath}
\usepackage{bbm}
\usepackage{enumitem}

\newcommand{\ket}[1]{| #1 \rangle}

\newcommand{\tr}[0]{ {\rm tr}}

\newcommand{\half}[1]{{ \rm  h}}
\newcommand{\Oorderof}{\mathcal{O}}
\newcommand{\orderof}[1]{\Oorderof(#1)} 

\newcommand{\for}[0]{\quad  \textrm{for} \quad}

\newcommand{\dist}{{\rm dist}}

\newcommand{\co}{{\rm c}}

\newcommand{\const}{{\rm const}}
\newcommand{\EqDef}{:=}
\newcommand{\diam}{{\rm diam}}
\newcommand{\LG}{\mathcal{G}}
\newcommand{\Vn}{\mathcal{V}}

\newcommand{\flo}{{\rm F}}

\def\beq{\begin{equation}}
\def\eeq{\end{equation}}
\def\nbeq{\begin{equation*}}
\def\neeq{\end{equation*}}
\def\<{\langle}
\def\>{\rangle}

\def\tr{{\rm tr}}

\theoremstyle{definition}
\newtheorem{theorem}{Theorem}
\newtheorem{lemma}{Lemma}
\newtheorem{corollary}{Corollary}

\usepackage{ulem}

\begin{document}

\title{Floquet-Magnus Theory and Generic Transient Dynamics in Periodically Driven Many-Body Quantum Systems}


\newcommand{\affA}{
WPI, Advanced Institute for Materials Research, Tohoku University, Sendai 980-8577, Japan
}

\newcommand{\affB}{
Department of Physics, Graduate School of Science, University of Tokyo, Bunkyo-ku, Tokyo 113-0033
}

\newcommand{\affC}{
Department of Physics, Keio University, 3-14-1 Hiyoshi, Kohoku-ku, Yokohama, Japan 223-8522
}

\author{Tomotaka Kuwahara}
\affiliation{\affA,\affB}

\author{Takashi Mori}
\affiliation{\affB}

\author{Keiji Saito}
\affiliation{\affC}

\begin{abstract}
This work explores a fundamental dynamical structure for a wide range of many-body quantum systems under periodic driving. Generically, in the thermodynamic limit, such systems are known to heat up to infinite temperature states in the long-time limit irrespective of dynamical details, which kills all the specific properties of the system. 
In the present study, instead of considering infinitely long-time scale, we aim to provide a general framework to understand the \textit{long but finite time} behavior, namely the transient dynamics. In our analysis, we focus on the Floquet-Magnus (FM) expansion that gives a formal expression of the effective Hamiltonian on the system.
Although in general the full series expansion is not convergent in the thermodynamics limit, we give a clear relationship between the FM expansion and the transient dynamics. 
More precisely, we rigorously show that  a truncated version of the FM expansion accurately describes the exact dynamics  for a certain time-scale.
Our theory reveals an experimental time-scale for which non-trivial dynamical phenomena can be reliably observed. 
We discuss several dynamical phenomena, such as the effect of small integrability breaking, efficient numerical simulation of periodically driven systems,  dynamical localization and thermalization. 
Especially on thermalization, we discuss a generic scenario on the prethermalization phenomenon in periodically driven systems.

\end{abstract}

\maketitle


\section{Introduction} \label{sec:introduction}
\subsection{Physical background}

Periodically driven quantum dynamics has recently attracted much attention in experimental as well as theoretical studies 
~\cite{bukov2014universal,RevModPhys.83.863,eisert2015quantum,grifoni1998driven,kohler2005driven} 
as it offers a promising way for exploring novel quantum phenomena which would be difficult or impossible to observe otherwise. 
Although the instantaneous Hamiltonian at each time step is very simple,
dynamical behavior can be highly nontrivial. Remarkable dynamical phenomena include 
dynamical localization~\cite{
PhysRevA.77.010101,
PhysRevLett.49.509,
PhysRevB.34.3625,
PhysRevB.89.165425,d2013many},
coherent destruction of tunneling~\cite{
PhysRevLett.67.516,
grossmann1992localization,
grifoni1998driven},
localization-delocalization transition~\cite{
PonteMBL2015,
abanin2014theory,
MBL_Analabha_2015,
ponte2015periodically,
PhysRevLett.115.030402},
and
dynamical phase transitions~\cite{
PhysRevLett.107.060403,
PhysRevA.86.063627,
shirai2014novel,
pineda2014two,
PhysRevLett.108.043003}.
%
Moreover, recent experimental development has rapidly opened new possibilities to 
control quantum systems under periodic driving, e.g., in the context of
quantum transport~\cite{
kohler2005driven,
PhysRevB.85.155425,
PhysRevLett.107.216601,
PhysRevB.84.235108},
quantum topological phases~\cite{
jotzu2014experimental,
PhysRevLett.112.156801,
lindner2011floquet,
PhysRevLett.113.266801,
PhysRevX.3.031005,
PhysRevX.4.031027,
PhysRevLett.108.225304} 
and detections of the Majorana Fermion~\cite{PhysRevLett.106.220402,PhysRevLett.111.136402} and the Higgs mode in condensed matter~\cite{matsunaga2014light}.

One of main subjects in driven quantum many-body systems is to understand the thermodynamical structure of steady states. With a few exceptions \cite{d2013many,PhysRevLett.107.060403,PhysRevA.86.063627,shirai2014novel,pineda2014two,PhysRevLett.108.043003}, recent studies mostly focus on driving simple (often non-interacting particles) Hamiltonians. 
However, the integrability-breaking terms unavoidably exist in the realistic experimental conditions.
When one looks at long-time behavior, even small integrability-breaking terms are relevant to the dynamics and cause significant effects on the final steady states. 
This provides the motivation to understand the long-time behavior of `generic' many-body systems under periodic driving.
Long-time behaviors in driven non-integrable systems are in general very complicated and cannot be captured with simple techniques such as the rotating-wave approximation and the transfer matrix technique of Landau-Zener transitions~\cite{PhysRevA.77.010101} etc., 
and hence one is obliged to rely on numerical calculations. 

Recently, true steady states in the long-time limit have been intensively studied using large scale numerical calculations~\cite{d2013many,Rigol_DAlessio2014,ponte2015periodically}. 
In the long-time limit, periodically driven many-body systems are in general expected to heat up to infinite temperature. 
This is a consequence from the analogy of eigenstate thermalization hypothesis (ETH) in non-driven many-body systems~\cite{rigol2008thermalization}. 
The ETH implies that each energy eigenstate of Hamiltonian is indistinguishable from the microcanonical ensemble with the same energy.
In periodically driven systems, the energy is no longer conserved and hence the extension of ETH to the driven case indicates that the steady state is a state of infinite temperature (i.e., the completely random state). 
Thus, due to the heating effect, any information reflected from the system is invisible in the \textit{infinite-time scale.}

However, the experiments on many-body quantum systems do not focus on the long-time limit; rather, they are interested in the transient dynamics for the \textit{experimental time scale.}
In this time scale, the heating process cannot necessarily occur, and hence a critical task that follows is to clarify the time-scale during which one can observe {\it transient behavior} that can show nontrivial phenomena. 
So far, most studies on  transient dynamical properties are based on numerical calculations with phenomenological arguments.

In this paper, we establish a new framework that describes transient dynamics for a wide range of many-body systems. 
Our analysis is based on the Floquet-Magnus theory. 
We show that as long as we consider a finite-time scale, the time-evolution of the system is approximately governed by a well-defined effective Hamiltonian.
We will give a prescription to obtain the effective Hamiltonian and reveal the time scale for which the system is governed by this Hamiltonian.
With the technique, we also discuss several dynamical phenomena relevant to the transient dynamics 
including thermalization process in driven systems.


\subsection{Outline of the basic framework}

\begin{figure}
\centering
\includegraphics[clip, scale=0.5]{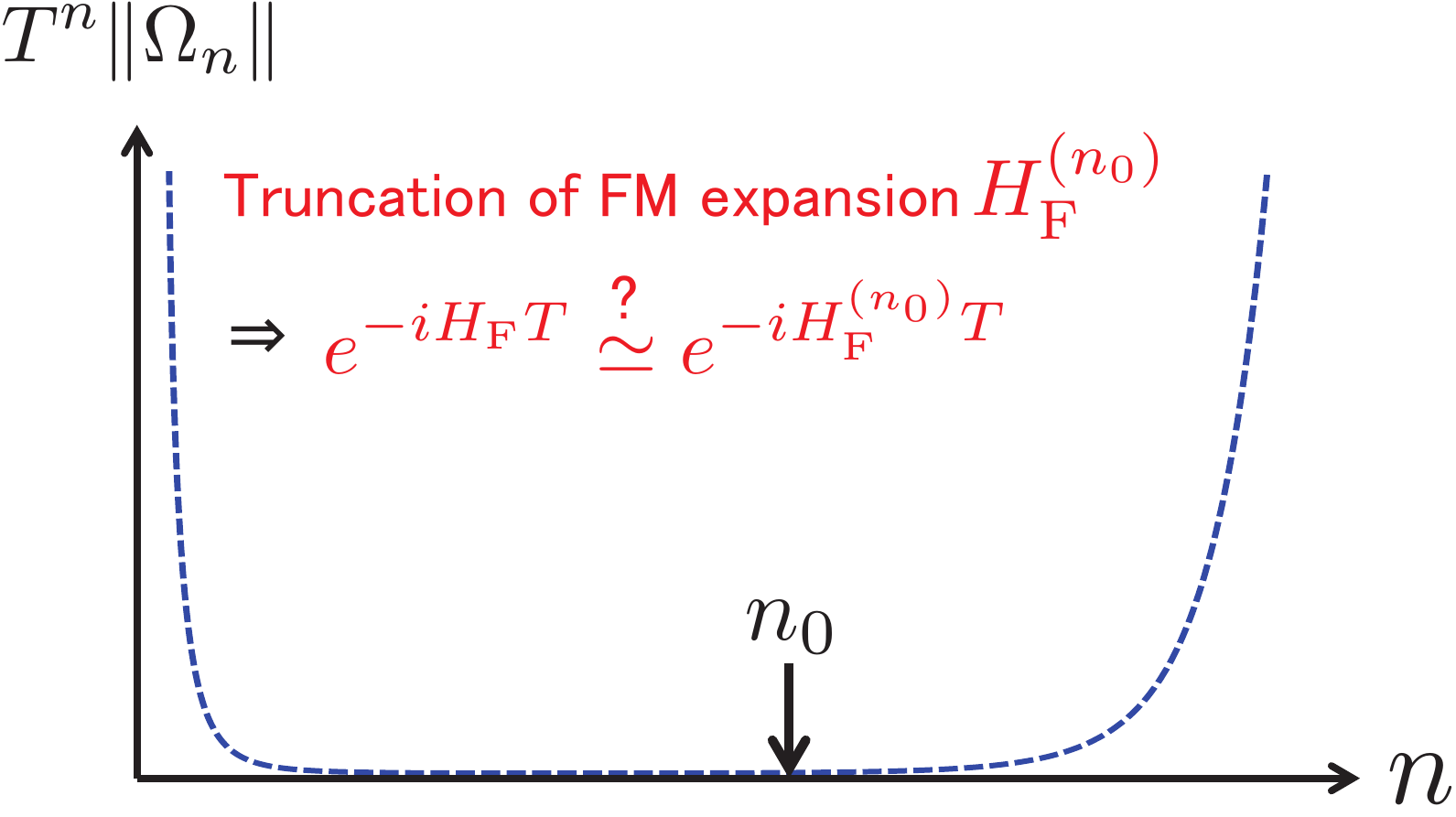}
\caption{In the thermodynamic limit, the FM-expansion~\eqref{eq:FM} diverges in general (See Fig.~\ref{fig:Magnus_numerical} for the numerical demonstration). Nevertheless, as we show, the truncated expression (\ref{eq:FM_finite}) at some order $n_0$ accurately describes the finite-time behavior of the periodic dynamics. 
}
\label{fig:FL_setup}
\end{figure}

In our analysis, the Floquet theory plays a fundamental role~\cite{PhysRev.138.B979} in defining the effective Hamiltonian that describes stroboscopic time-evolution for every period. This effective Hamiltonian is referred to as the \textit{Floquet Hamiltonian} 
$H_\flo$ and is defined as 
\begin{align}
e^{-i H_\flo T} \EqDef \mathcal {T} \bigl [e^{-i\int_{t_0}^{t_0+T} H(t) dt} \bigr] \notag
\end{align}
with $\mathcal {T}$ the time-ordering operator, where the Hamiltonian is periodic in time: $H(t)=H(t+T)$. Throughout the paper, we set $\hbar=1$ and $t_0=0$, and we do not consider the degree of freedom on the Floquet gauge~\cite{bukov2014universal,PhysRevX.4.031027,PhysRevLett.91.110404,PhysRevA.68.013820}. The Floquet Hamiltonian has full information on the thermodynamic properties of the driven systems, and hence finding this is one of the central problems in periodically driven systems. 
However, the time-ordering operation in the unitary time-evolution is, in general, difficult to analyze, and hence finding the Floquet Hamiltonian is a highly nontrivial problem
~\cite{
bukov2014universal,
eckardt2015consistent,
PhysRevLett.91.110404,
PhysRevA.68.013820,
PhysRevLett.111.175301,
PhysRevE.91.032923,PhysRevLett.115.075301}.


The Floquet-Magnus (FM) expansion~\cite{magnus1954exponential,blanes2009magnus} is known to give a formal expression of the Floquet Hamiltonian as follows
\begin{align}
 H_\flo &= \sum_{n=0}^\infty T^n \Omega_n \quad (\textrm{FM expansion}) \, , \label{eq:FM} 
\end{align}
where explicit forms of the terms $\{\Omega_n\}_{n=0}^\infty$ are given by Eq.~\eqref{eq:Omega_n} in Subsection~\ref{FL:Floquet-Magnus Expansion}. In this paper, we also consider the truncated expansion up to the $n$-th order, which is defined as
\begin{align}
H_\flo ^{(n)}:=\sum_{m=0}^nT^m\Omega_m \quad (\textrm{Truncated FM expansion}) .
\label{eq:FM_finite}
\end{align}
The FM expansion (\ref{eq:FM}) is useful especially for the high-frequency limit in finite-size systems, where the higher-order contribution is negligible. On the other hand, in
the case of finite frequencies, the higher-order contribution should always be considered. 
However,  it has been recognized that, in general, the FM expansion is not convergent series expansion~\cite{bukov2014universal,blanes2009magnus}. 
The divergence problem in the FM expansion has been a long-standing obstacle in analyzing periodically driven systems, and hence 
our understanding of the Floquet Hamiltonian for many-body systems is so far extremely limited~\cite{d2013many,Rigol_DAlessio2014,ponte2015periodically}.

The divergence of the FM expansion is expected to have an interesting physical meaning \cite{d2013many,Rigol_DAlessio2014,ponte2015periodically,MBL_Analabha_2015,PonteMBL2015,PhysRevLett.115.030402}. 
In periodically driven systems, even the energy is not a conserved quantity. Hence,
the ergodic time-evolution implies that the entire phase space can be covered by the time evolution from an arbitrary initial state. References~\cite{Rigol_DAlessio2014,PonteMBL2015} have numerically shown the signature of ergodic dynamics in the time-evolution operator showing that it is described by a random matrix. 
Based on these observations, one anticipates a deep connection between the divergence of the FM expansion and the quantum ergodicity~\cite{d2013many,Rigol_DAlessio2014,ponte2015periodically,MBL_Analabha_2015,PonteMBL2015,PhysRevLett.115.030402}.
This connection indicates that the divergence is usually unavoidable because
generic non-integrable many-body systems are believed to be ergodic in the thermodynamic limit~\cite{PhysRevLett.108.110601,PhysRevA.43.2046,Rigol_DAlessio2014,PhysRevE.60.3949,PhysRevE.50.888,rigol2008thermalization,PhysRevE.79.051129}.

Even though the FM expansion is divergent in general, we empirically expect that the finite truncation of the FM expansion (\ref{eq:FM_finite}) can give useful information on the transient dynamics. 
We will show that the FM expansion has an optimum order to approximate the transient dynamics \textit{even if it is divergent in the limit of $n\to \infty$}; the figure~\ref{fig:FL_setup} schematically shows the typical behavior of the FM expansion. 
The validity of the finite truncation of the FM expansion has been studied in several specific cases such as the Friedrichs model~\cite{Mori_Friedrichs} and the NMR of solids~\cite{PhysRevB.25.6622}. Our purpose in this paper is to generalize these results and to give a rigorous relationship between the FM expansion and general properties of transient quantum dynamics.

\begin{table}[b]
  \caption{Effective Hamiltonian for different time scales}
  \label{tab:FL} 
\begin{ruledtabular}
\begin{tabular}{lr}
\textrm{\textbf{Time-scale}}&
\textrm{\textbf{Effective Hamiltonian}}\\
\colrule
  finite  ($t\lesssim e^{\orderof{\omega}}$) & $H_\flo^{(n_0)}$ ($n_0=\orderof{\omega}$)  \\
   infinite ($t=\infty$)    &  random matrix  \\
\end{tabular}
\end{ruledtabular}
\end{table}

In the present study, we consider a generic isolated quantum system on a lattice,
where we do not impose any assumption such as the integrability or symmetry conditions. 
We only assume that the interaction couplings between spins (particles) are few-body
in Hamiltonian.
This class of Hamiltonians has, thus far, been employed to study on quantum information oriented problems~\cite{arad2014connecting,PhysRevB.85.195145,arad2013area,kuwahara2015local,kuwahara2015locality}. 
Here, we apply the techniques developed for such problems to the Floquet theory, and rigorously show the condition 
under which the truncated expression (\ref{eq:FM_finite}) can be used and estimate its accuracy, towards the main aim of this study.
As shown in Table~\ref{tab:FL}, we will show that by choosing the truncated Hamiltonian $H_\flo^{(n_0)}$ with $n_0=\orderof{\omega}$, the error of $\|e^{-iH_\flo T} -e^{-iH_\flo^{(n_0)}T}\|$ decays exponentially as $\omega$ increases.
This fact leads to the notion of \textit{quasi-stationary} states for long but finite time-scale; 
the system first relaxes to an intermediate state which is characterized by the effective Hamiltonian $H_\flo^{(n_0)}$, 
and then reaches the final steady state (i.e., the infinite temperature state) very slowly. 
We will see that such a quasi-stationary state is maintained up to an exponentially long time scale of $e^{\orderof{\omega}}$.


The  present paper is organized as follows.
In Sec.~\ref{FL:Setup and main results}, we show the setup of systems and our main results on the generic properties of the transient dynamics.
As shown in Sec.~\ref{FL:Relations to known results}, our results are related to various facts in fundamental physics.
In Sec.~\ref{FL:Summary}, we summarize our results and mention areas for further development. 
Finally, Section~\ref{FL:Proof of our results} is devoted to proving the main theorems.

\section{Setup and main results} \label{FL:Setup and main results}

In this section, we first introduce the setup of our system and show the
convergence of the Floquet-Magnus expansion up to a finite order with several numerical demonstrations.
We then give our main theorems on the transient dynamics.

\subsection{Generic few-body Hamiltonians} \label{FL:Setup}

We consider a spin system of finite volume with each spin having a $d$-dimensional Hilbert space, and we label each spin by $i=1,2,\ldots  N$. 
We denote the set of all spins by $\Lambda=\{1,2,\ldots  N\}$, a partial set of sites by $X$, and the cardinality of $X$, that is, the number of sites contained in $X$, by $|X|$ (e.g. $X=\{i_1,i_2,\ldots, i_{|X|}\}$).
 We also denote the complementary subsets of $X$ by $X^\co$; that is, $X\oplus X^\co=\Lambda$.
Note that the spins are not necessarily located next to each other on the lattice.

We consider a general periodically driven system on an arbitrary lattice,
the Hamiltonian of which is given by
$
H(t) = H_0 + V (t),\notag 
$
where $H_0$ is the static Hamiltonian without the driving potential and $V (t) = V (t+T)$ denotes the driving Hamiltonian with the period $T$ (with the frequency $\omega:=1/T$). 
Throughout the paper, we consider the system which is governed by a generic few-body Hamiltonian, that is, 
the Hamiltonian contains at most $k$-body interactions with finite $k$:
 \begin{align}
H_0=\sum_{|X|\le k} h_X ,\quad V(t)=\sum_{|X|\le k} v_X(t), 
\label{few_body_Hamiltonian_eq}
\end{align}
where $h_X$ is an operator acting on a spin subset $X$.
More explicitly, it can be given in the form of
 \begin{align}
H=&\sum_i h_i s_i + \sum_{i_1,i_2} J_{i_1,i_2} s_{i_1} s_{i_2} +\sum_{i_1,i_2,i_3} J_{i_1,i_2,i_3} s_{i_1} s_{i_2} s_{i_3} \notag \\
&+\cdots +\sum_{i_1,i_2,\ldots,i_k} J_{i_1,i_2,\ldots,i_k} s_{i_1} s_{i_2} \cdots s_{i_k}, \notag 
\end{align}
where $\{s_i\}$ are operator bases on the $i$th spin; for example, it can be given by the Pauli matrices for $(1/2)$-spin systems, namely $\{s_i\}=\{\sigma_i^x,\sigma_i^y,\sigma_i^z\}$.
This definition of the Hamiltonian encompasses  almost all interesting quantum many-body systems
with short-range interactions such as the the $XY$ model~\cite{ref:XY}, the Heisenberg model~\cite{ref:Haldane1983-HeisenbergI,ref:Haldane1983-HeisenbergII} and the AKLT model~\cite{ref:AKLT}, as well as models with long-range interactions such as the Lipkin-Meshcov-Glick model~\cite{ref:LMG}.
Typically, we have $k=2$ (i.e., two-body interaction), but several exceptions exist such as the cluster-Ising model~\cite{PhysRevB.88.125117} ($k=3$), the toric code model~\cite{kitaev2003fault} ($k=4$) and the string-net model~\cite{PhysRevB.71.045110} ($k=6$).
Although we do not treat fermionic systems explicitly, our discussions can be also applied to
local fermionic systems because they can be mapped into local spin systems~\cite{PhysRevLett.95.176407,1742-5468-2005-09-P09012}.

We then introduce a parameter $J$ as a local interaction strength (or one-particle energy) of the system:
 \begin{align}
\sum_{X: X\ni i}( \|h_X\| +\|v_X(t)\|) \le J  \for \forall i \in \Lambda,
\label{eq:extensiveness}
\end{align}
where $\|\cdots\|$ is the operator norm and $\sum_{X:X\ni {i}} $ denotes the summation with respect to the supports containing the spin $i$.
Note that we thereby obtain $\sum_{X: X\ni i} \|h_X\| \le J$ and $\sum_{X: X\ni i} \|v_X(t)\| \le J $.
This condition indicates that the energy change due to one spin is bounded by a finite value $J$.
Throughout the paper, we use the notations
 \begin{align} 
&V_0:=\sum_{|X|\leq k} \frac{1}{T}\int_0^T \|v_X(t)\|dt , \quad \lambda := 2kJ,\notag \\
&\orderof{\omega} = \const \cdot \frac{\omega}{\lambda} \quad {\rm and} \quad  \orderof{T} = \const \cdot \lambda T ,\label{Notations_V_0_lambda}
 \end{align}
where $V_0T$ denotes the driving amplitude in one period.
In our theory, the parameter $\lambda$ characterizes typical properties of the system.

\subsection{Floquet-Magnus (FM) expansion}  \label{FL:Floquet-Magnus Expansion}

A possible method to calculate the Floquet Hamiltonian $H_\flo$ is to expand it with respect to the period $T$ as in Eq.~\eqref{eq:FM}, namely $H_\flo = \sum_{n=0}^\infty T^n \Omega_n$.
The FM expansion gives each of the terms $\{\Omega_n\}_{n=0}^\infty$ as follows~\cite{bialynicki1969explicit}:
\begin{widetext}
\begin{align}
\Omega_n =
&\frac{1}{(n+1)^2} \sum_{\sigma}
(-1)^{n-\theta(\sigma)} \frac{\theta(\sigma)!(n-\theta(\sigma))! }{n! }
 \notag \\
&\times \frac{1}{i^n T^{n+1}} \int_0^Tdt_{n+1}\dots \int_0^{t_3}dt_2 \int_0^{t_2}dt_1
\left[ H(t_{\sigma(n+1)}), \left[ H(t_{\sigma(n)}),\ldots,\left[ H(t_{\sigma(2)}),H(t_{\sigma(1)})
\right]\ldots\right]\right] , ~~~~
\label{eq:Omega_n}
\end{align}
where $\sigma$ is the permutation and $ \theta(\sigma):=\sum_{i=1}^n\theta(\sigma(i+1)-\sigma(i)), $ with $\theta(\cdot)$ the usual step function. 
For example, the first three terms in the expansion read (See~\cite{Magnus_higher} for higher-order terms)
\begin{align}
&\Omega_0 (T) =\frac{1}{T} \int_{0}^{T} H(t_1) dt_1, \quad \Omega_1 (T) =\frac{1}{2iT^2} \int_{0}^{T}dt_1\int_{0}^{t_1}  dt_2  [H(t_1),H(t_2)]   ,\notag  \\
&\Omega_2 (T) =-\frac{1}{6T^2} \int_{0}^{T}dt_1\int_{0}^{t_1}  dt_2\int_{0}^{t_2}  dt_3 \bigl(  [H(t_1),[H(t_2),H(t_3)]]  +[H(t_3),[H(t_2),H(t_1)]] \bigr). \notag 
\end{align}
\end{widetext}
Without loss of generality, we can put $\int_0^TV(t)dt=0$ and, thus, $\Omega_0=H_0$.
The FM expansion is considered a useful tool to treat a periodically driven system when the period $T$ of the driving is sufficiently small.

From a general discussion, we can evaluate the upper bound of $\|T^n\Omega_n\|$ as
\begin{align}
\|T^n\Omega_n\| \le \biggl(\frac{1}{\xi} \int_0^T\|H(t)\|dt \biggr)^n,\notag 
\end{align}
where $\xi$ is a universal constant~\cite{blanes2009magnus,pechukas1966exponential,karasev1976infinite,blanes1998magnus,moan2008convergence}.
Then, the convergence of the FM expansion is ensured only for the case of
$
\int_0^T\|H(t)\|dt\leq \xi ,
$
and it is not satisfied  for a macroscopic system $(\|H(t)\|\propto N)$, unless the period $T$  scales with the total system size $N$.
Thus, this simple estimation indicates the divergence in the thermodynamic limit.

When the Hamiltonian is given by a few-body operator~\eqref{few_body_Hamiltonian_eq} with the condition \eqref{eq:extensiveness}, much stronger bound for $\Omega_n$ exists:

\begin{lemma}\label{FL:FM_bound}
For each of the terms $\Omega_n$ with $n\ge1$, 
\begin{align}
\|\Omega_n\|\leq\frac{2V_0 \lambda^n}{(n+1)^2}n!=:\overline{\Omega}_n,
\label{eq:norm_Omega}
\end{align}
where $\lambda$ and $V_0$ are defined in Eq.~\eqref{Notations_V_0_lambda}.
The proof is given in Appendix~\ref{FL:proof_FM_bound}.
\end{lemma}

The bound of Eq.~(\ref{eq:norm_Omega}) implies that the FM expansion is convergent up to $n\approx(\lambda T)^{-1}$ in the sense of
$
\|\Omega_n\|T^n>\|\Omega_{n+1}\|T^{n+1}
$
for $n\lesssim (\lambda T)^{-1}$.
However, for $n\gtrsim (\lambda T)^{-1}$, the FM expansion begins to diverge.
In the next subsection, we show that even if the FM expansion might diverge for $n\to \infty$, the truncation~\eqref{eq:FM_finite}
 characterizes the quantum dynamics with great accuracy.

%


 \begin{figure}
\centering
\subfigure[Logarithmic plot of $T^n \|\Omega_n\|$]{
\includegraphics[clip, scale=0.195]{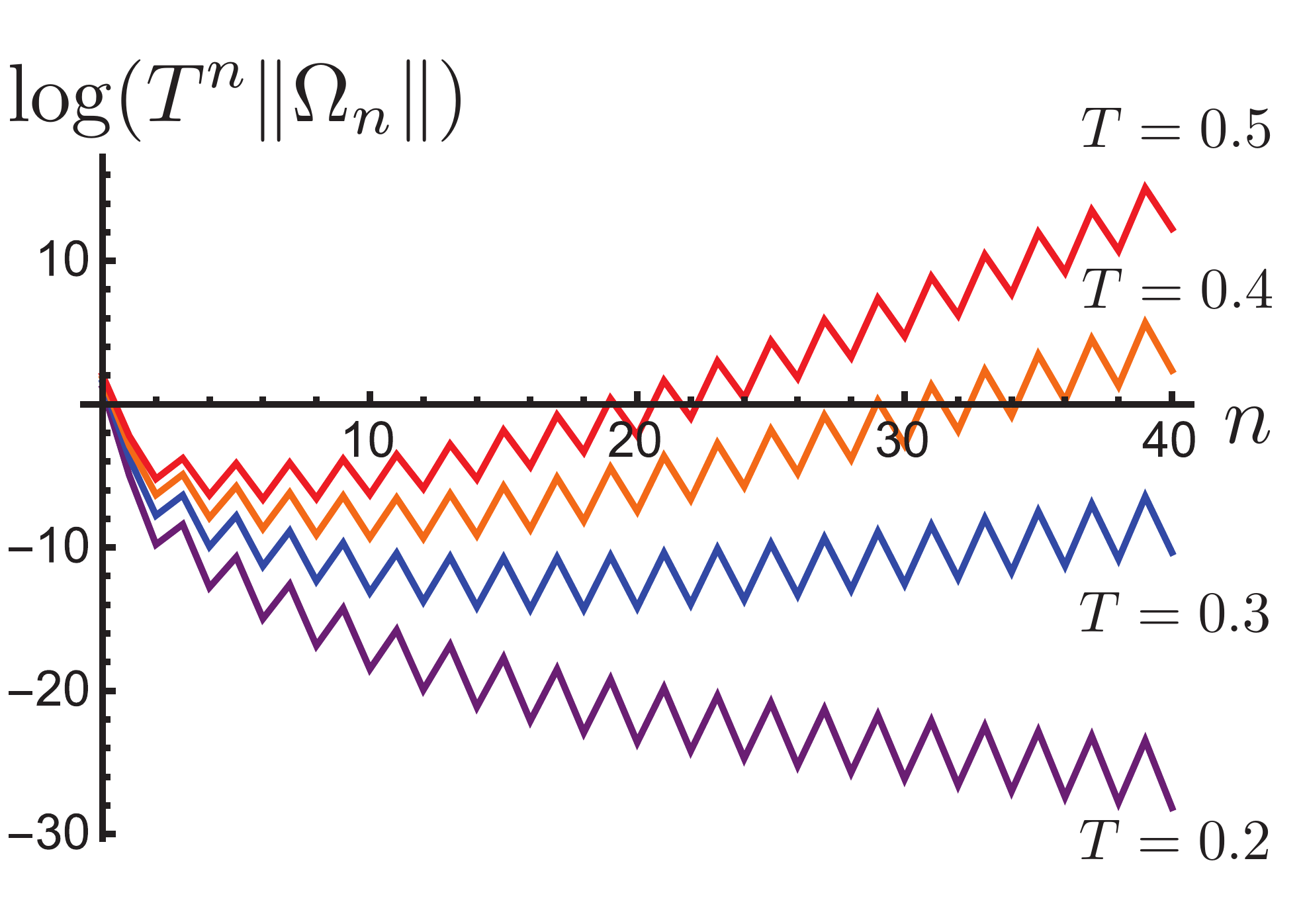}
}
\subfigure[Logarithmic plot of $\|H_\flo^{(n)}\|$]{
\includegraphics[clip, scale=0.195]{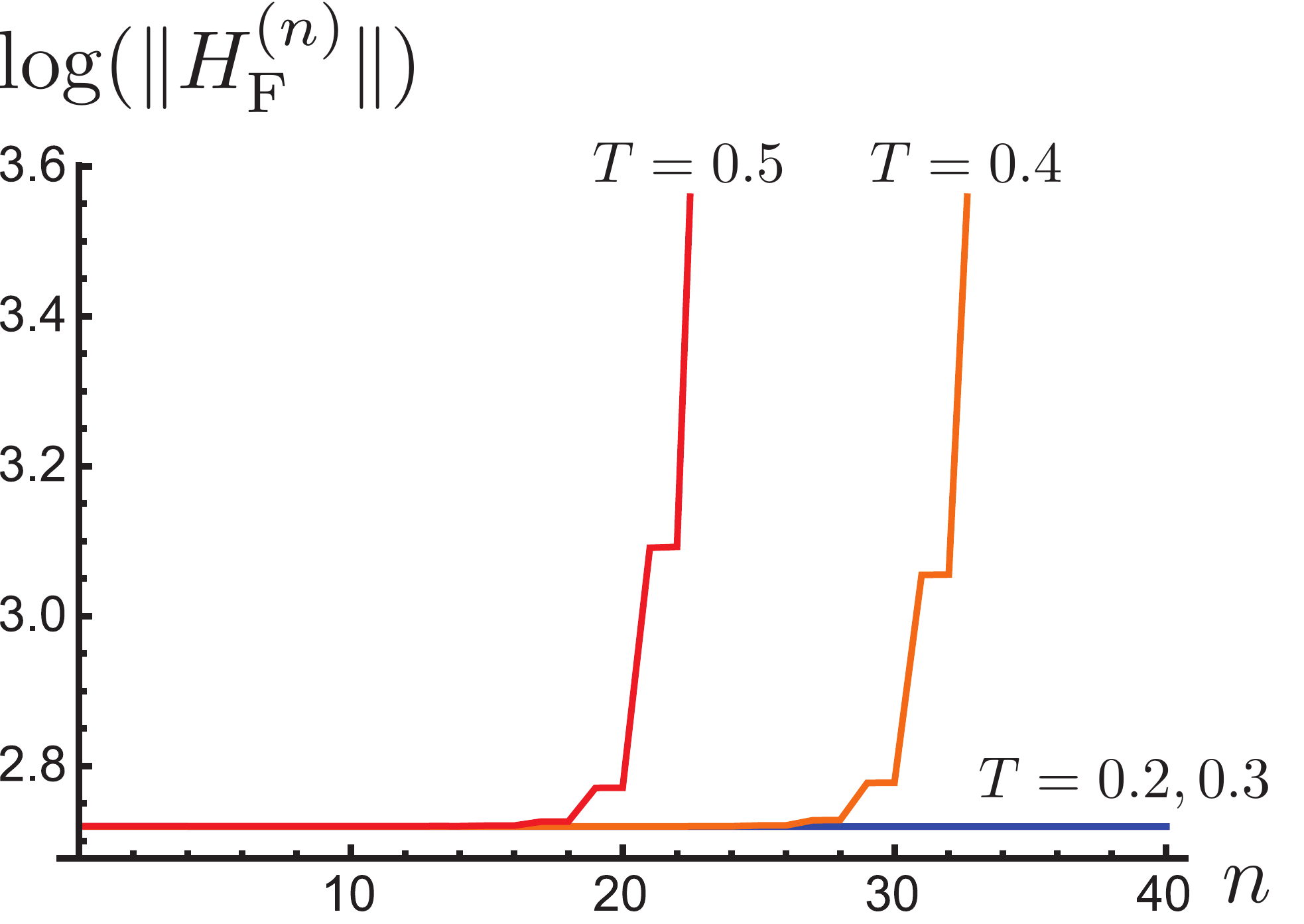}
}
\subfigure[Logarithmic plot of $\|e^{-i H_{\flo} T} - e^{-i H_{\flo}^{(n)} T}\|$]{
\includegraphics[clip, scale=0.35]{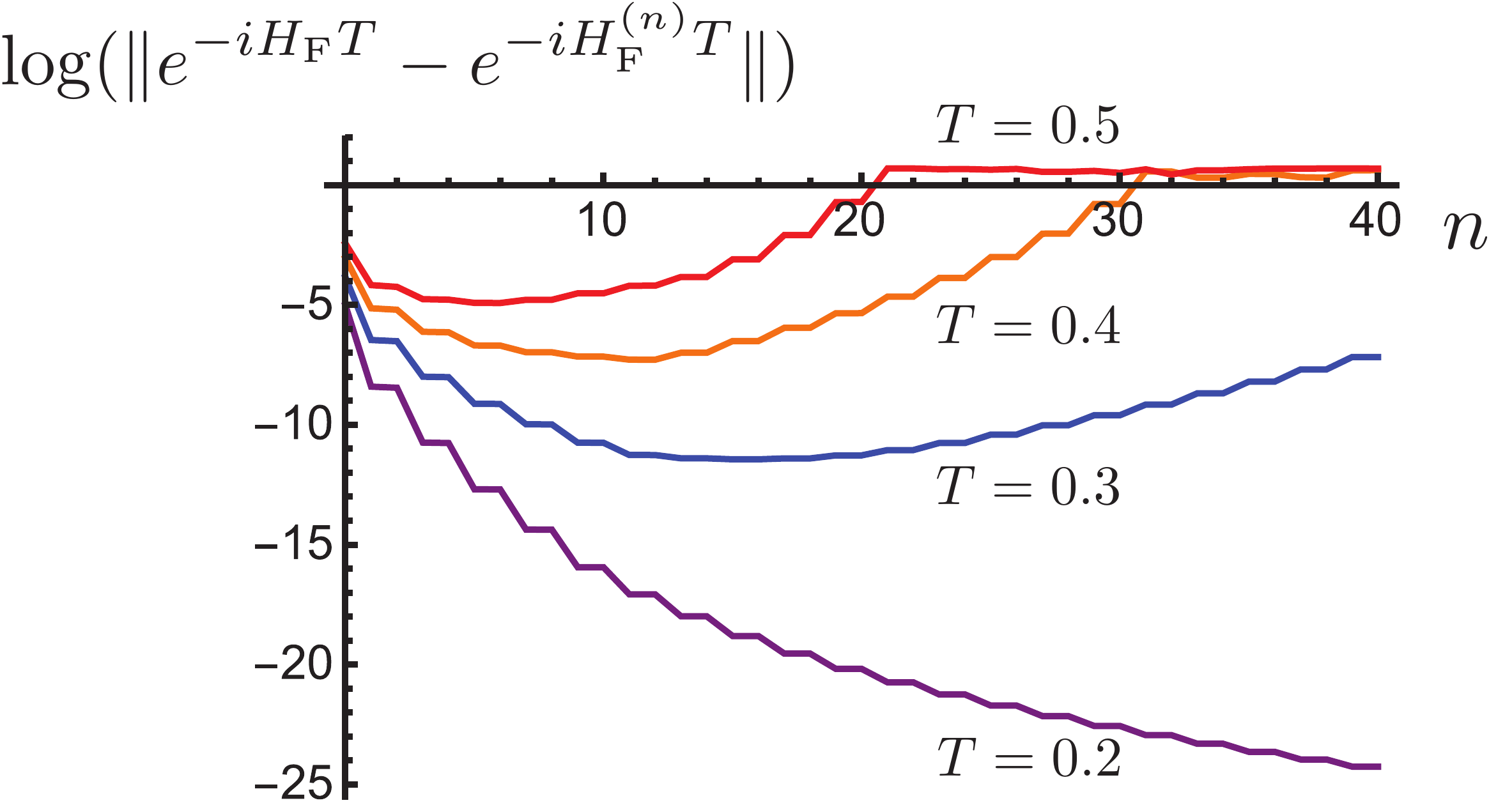}
}
\caption{The logarithmic plots of (a) norms of each term $T^n \Omega_n$ in FM expansion, (b) norms of the truncated Floquet Hamiltonians $H_{\flo}^{(n)}$, (c) norms of $e^{-i H_{\flo} T} - e^{-i H_{\flo}^{(n)} T}$ with respect to the order $n$.
We consider the Heisenberg model with a time-periodic field for three cases of the period $T=0.2,0.3,0.4,0.5$, which correspond to the purple, blue, orange, and red curves, respectively.
}
\label{fig:Magnus_numerical}
\end{figure}

\subsection{Numerical demonstration in anisotropic Heisenberg chain}
Before showing our main theorems, we would explain our results in a visual way. 
For the purpose, in Fig.~\ref{fig:Magnus_numerical}, we give numerical calculations of the FM expansion.
Here, we consider an 8-spin anisotropic Heisenberg chain with a time-periodic field: 
\begin{align}
H(t) = \sum_{i=1}^8 \left(\frac{3}{2}\sigma_i^x\sigma_{i+1}^x+\sigma_i^y\sigma_{i+1}^y+\frac{1}{2}\sigma_i^z\sigma_{i+1}^z+t \sigma_i^z\right) ,
\notag
\end{align}
for $0 < t \le T$ with $H(t)=H(t+T)$,
where $\{\sigma_i^\xi\}_{\xi=x,y,z}$ are the Pauli matrices and we assume the periodic boundary condition, namely $\sigma_{9}=\sigma_1$. 
We calculate the FM expansion for four cases of $T=0.2, 0.3, 0.4, 0.5$ up to the order $n= 40$.

We plot the norms of the expansion terms $T^n \Omega_n$ (Fig.~\ref{fig:Magnus_numerical} (a)) and the truncated Floquet Hamiltonians $H_{\flo}^{(n)}$ (Fig.~\ref{fig:Magnus_numerical} (b)).
The calculation implies that the FM expansion is neither conditionally convergent nor absolutely convergent.
Thus, for $T\gtrsim 0.3$, the Magnus expansion completely breaks down for $n\gg1$, while the Magnus expansion converges up to a small $n$. 
This is consistent to the Lemma~\ref{FL:FM_bound}.
Relevant to them, we demonstrate the accuracy of the truncated FM expansion in Fig.~\ref{fig:Magnus_numerical} (c), 
where we plot the error $\|e^{-iH_\flo T}  - e^{-iH_\flo^{(n)}T}\|$ with respect to the truncated order $n$. 
We can see that the qualitative behavior is synchronized with that of $\| T^n\Omega_n\|$ in Fig.~\ref{fig:Magnus_numerical} (a).
In the following, we show the analytical reasonings to these observations.


\subsection{Transient dynamics for generic few-body Hamiltonians}  \label{FL:Validity of the truncated Floquet Hamiltonian}

We shall prove the following theorem on the validity of the truncated Floquet Hamiltonian for the time evolution:
\begin{theorem}\label{FL:Metastability}
Consider $H_0$ and $V(t)=V(t+T)$ be few-body operators with the local interaction strength $J$, respectively, and
assume that $T$ satisfies 
$
T\le 1/(4\lambda) . 
$
Then, the time evolution under the Hamiltonian $H(t)=H_0+V(t)$ is close to that under the truncated Floquet Hamiltonian $H_\flo ^{(n_0)}=\sum_{m=0}^{n_0}\Omega_mT^m$ with 
\begin{align}
n_0:=\left \lfloor\frac{1}{16\lambda T} \right\rfloor = \orderof{\omega} \notag 
\end{align}
 in the sense that
\begin{align}
\|e^{-iH_\flo T}-e^{-iH_\flo^{(n_0)}  T} \| \le 6V_0 T  2^{-n_0} ,
\label{eq:theorem}
\end{align}
where $\lfloor\cdot \rfloor $ is the floor function.
See Eq.~\eqref{Notations_V_0_lambda} for the definition of $\lambda$ and $V_0$.
We give the proof in Section~\ref{FL:Proof of our results_thm1}.
\end{theorem}

Note that the right hand side of \eqref{eq:theorem} is exponentially small for $\omega$, as $V_0 T e^{-\orderof{\omega}}$. 
Although the above theorem compares the time evolutions only for one period, it immediately follows that for $t=mT$ with $m$ a positive integer:
\begin{align}
\| e^{-iH_\flo  mT}-e^{-iH_\flo^{(n_0)} mT}\| \leq 6V_0   mT 2^{-n_0} .  \notag 
\end{align}
Thus, this theorem indicates that the exact time evolution is well approximated by the time evolution under the $n_0$-th order truncated Floquet Hamiltonian up to time $t\approx e^{\orderof{\omega}}$.
Even if the Floquet-Magnus expansion is not convergent and the system might eventually heat up to infinite temperature, the transient dynamics of the system is governed by the truncated Floquet Hamiltonian for an exponentially long time with respect to the frequency of external driving.

In practice, we are often interested in the dynamical behavior by $H_\flo^{(n)}$ with $n$ less than $n_0$; indeed, we mostly treat the case of 
$n=0$, namely the average Hamiltonian.
By using Eq.~(\ref{eq:norm_Omega}), we can also know how close the exact time evolution is to that under the $n$-th order truncated Hamiltonian:
\begin{corollary}\label{FL:m_period_theorem}
Under the conditions in Theorem~\ref{FL:Metastability}, we have 
\begin{align}
&\| e^{-iH_\flo  T}-e^{-iH_\flo^{(n)} T} \|
\leq  6V_0 T  2^{-n_0} + \overline{\Omega}_{n+1}T^{n+2}
\label{eq:general_n}
\end{align}
for an arbitrary $n\leq n_0$, where $\overline{\Omega}_n$ is defined in Eq.~\eqref{eq:norm_Omega}.
We show the proof in Appendix~\ref{Proof of Corollary1}.
\end{corollary}
This corollary rigorously shows that the Floquet-Magnus expansion is, at least, an asymptotic expansion.


\subsection{Transient dynamics for short-range few-body Hamiltonians}  \label{FL:Validity of the truncated Floquet Hamiltonian}

In Theorem~\ref{FL:Metastability}, we have shown that the total unitary operator $e^{-iH_\flo T}$ is approximately described by $e^{-iH_\flo^{(n_0)} T}$.
However, the inequality~\eqref{eq:theorem} is meaningful only when $V_0 T  2^{-n_0} \lesssim 1$, 
or $\omega \gtrsim \log V_0$; 
because $V_0$ is proportional to the number of spins subjected to driving, 
the condition is not satisfied for finite frequencies under global driving ($V_0\propto N$) in the thermodynamic limit of $N\to \infty$.
In the following, we improve the condition by imposing a stronger restriction to the Hamiltonian, namely the assumption of the short-range interaction.

We here focus on the time evolution of a local region $L$ instead of the total system $\Lambda$; let 
$\rho$ be an arbitrary initial state and consider its reduced density matrix in the region $L$, namely 
$
\rho_L = \tr_{L^{\co}}(\rho), 
$
where $\tr_{L^{\co}}(\cdots)$ denotes the partial trace operation with respect to the spins in $L^{\co}$.
We now define 
\begin{align}
&\rho_L(mT ) = \tr_{L^{\co}}( e^{-iH_\flo T} \rho e^{iH_\flo T} ), \notag \\
&\rho_L^{(n_0)}(mT ) = \tr_{L^{\co}}( e^{-iH_\flo^{(n_0)} T} \rho  e^{iH_\flo^{(n_0)} T}  ) \label{def_rho_mT_2}
\end{align}
and obtain the upper bound of $\| \rho_L(mT ) -\rho_L^{(n_0)}(mT)\|_1$ with $\|\cdot\|_1$ the trace norm. 
Under the condition of the short-range interaction, 
the error between  $\rho_L(mT )$ and $\rho_L^{(n_0)}(mT)$ is exponentially small for $\omega$ 
up to $t \approx e^{\orderof{\omega}}$ as long as $|L| \lesssim e^{\orderof{\omega}}$.

In considering short-range interacting systems, we have to define the structure of the system explicitly (e.g., the square lattice)~\cite{ref:Hastings2006-ExpDec}.
We now define a set of the bonds $\Lambda_b$, i.e. pairs of spins $\{i_1,i_2\}$, $\{i_2,i_3\}$, $\{i_2,i_5\}$ and so on.
The form of $\Lambda_b$ decides the structure of the lattice. 
Based on this definition, we define the distance $\dist(X, Y)$ as the shortest-path length which one needs to connect the two partial sets $X$ and $Y$.

In this subsection, we introduce the following additional assumption to the Hamiltonian: 
 \begin{align}
\sum_{X: X \ni i,  \diam (X) \ge r } \| h_X \| \le  F(r) \for \forall i \in \Lambda , \notag 
\end{align}
where $\diam (X):= \sup_{\{i,j\} \in X} \dist(i,j)$ and the function $F(r)$ determines how the interactions decays as the spatial distance increases.
In this case, we can prove the Lieb-Robinson bound for arbitrary operators $O_X$ and $O_Y$~\cite{ref:Hastings2006-ExpDec,ref:LR-bound72,PhysRevLett.97.050401,ref:Nachtergaele2006-LR,PhysRevLett.114.157201}:
\begin{align}
\|[O_X(t),O_Y]\| \le \LG(l,t)\min (|X|,|Y|) \|O_X\| \cdot \|O_Y\| , \label{Lieb_Robinson_bound_G}
\end{align}
where $l=\dist(X,Y)$ and $\LG(l,t)$ is characterized by the form of the interaction decay $F(r)$.
Note that $\LG(l,t)$ is a monotonically increasing function with respect to $t$.
The Lieb-Robinson bound characterizes the non-locality of operator due to the time-evolution; that is,  
we can know how fast an operator $O_X$ spreads away from the region $X$ after a short time. 

Using the above notations, we can obtain the following theorem for short-range interacting Hamiltonians:
\begin{theorem}\label{FL:Metastability4}
Let us consider a $D$-dimensional system which satisfies \eqref{Lieb_Robinson_bound_G}.
Then, the exact time evolution of $\rho_L(m T)$ by $e^{-iH_\flo mT} $ is close to the approximate time evolution $\rho_L^{(n_0)}(mT)$ by $e^{-iH_\flo^{(n_0)} mT} $ with $n_0=\left \lfloor\frac{1}{16\lambda T} \right\rfloor$:
\begin{align}
&\| \rho_L(m T) - \rho_L^{(n_0)}(m T) \|_1  \notag \\
\le& 12J |L| mT 2^{-n_0/2 } + 2  |L| m  \LG(l_0,mT) \label{eq:theorem4}
\end{align}
with $l_0= \const \cdot 2^{n_0/(2D)}=e^{\orderof{\omega/D}}$,
where $\|\cdot\|_1$ denotes the trace norm.
We give the proof in Section~\ref{FL:Proof of our results_thm2}.
\end{theorem}

For example, in the case where the interaction decays exponentially ($F(r)\sim e^{-r}$), we know~\cite{ref:Hastings2006-ExpDec,ref:LR-bound72,PhysRevLett.97.050401,ref:Nachtergaele2006-LR}
\begin{align}
\LG(l,t) =c e^{- (l-v t)/\xi }\label{LR_bound_exp_short_range}
\end{align}
with $c$, $v$ and $\xi$ some constants depending on $\lambda$ and $F(r)$.
Then, we have $\LG(l_0,mT) =c \exp\bigl[- (e^{\orderof{\omega/D}}-v mT)/\xi \bigr] $.
Hence, by applying the Lieb-Robinson bound~\eqref{LR_bound_exp_short_range} to the inequality~\eqref{eq:theorem4}, 
the second term is negligibly small in comparison with the first term as long as  $mT \lesssim e^{\orderof{\omega}}$. 
Thus, in this time-scale, the error between $\rho_L(m T)$ and $\rho_L^{(n_0)}(m T)$ is as small as $|L|e^{-\orderof{\omega}}$.

On the other hand, when the interaction decays polynomially $(F(r) \sim 1/r^{\alpha})$, the Lieb-Robinson bound is given by~\cite{PhysRevLett.114.157201,ref:Hastings2006-ExpDec} 
\begin{align}
&\LG(l,t) =c  \frac{e^{v t}}{l^{\alpha}} \for D< \alpha \le 2D, \notag \\
&\LG(l,t) =c  e^{- (l/t^\gamma-v t)/\xi } + c'  \frac{t^{\alpha(1+\gamma)}}{l^{\alpha}}   \for \alpha > 2D \notag 
\end{align}
with $\{c, c', v, \xi,\gamma\}$ constants depending on $\lambda$ and $F(r)$.
Thus, from the inequality~\eqref{eq:theorem4}, in the former case, the error $\| \rho_L(m T) - \rho_L^{(n_0)}(m T) \|_1$ is exponentially small $e^{-\orderof{\omega}}$ only for the time $t \approx \orderof{\omega}$, while in the latter case, the error is exponentially small up to the time $t\approx e^{\orderof{\omega}}$.
This way, if the interaction decays rapidly enough $(\alpha >2D)$, we can ensure that the truncated FM expansion gives a good approximation as long as we look at the local region of the system ($|L| \lesssim e^{\orderof{\omega}}$). 
We emphasize that Theorem~\ref{FL:Metastability4} does not depend on the driving amplitude $V_0$ and is meaningful even for 
global driving ($V_0\propto N$) in the thermodynamic limit of $N\to \infty$.

\subsection{Probability of energy absorption}  \label{sec_FL:Heat_absorption}

We here discuss an energy absorption by the periodic driving. 
Let $\Pi_{\ge E}^{(n)}$ be a projection operator onto the eigenspaces of $H_\flo^{(n)}$ ($n\le n_0$) with energies larger than $E$. 
Then, for an arbitrary initial state $\ket{\psi_{\le E}}$ which is in a superposition of energies below $E$ (see Fig.~\ref{fig:FL_Energy_absorption}),
we define the probability to absorb energy $\Delta E$ after $m$ periods as
$
\| \Pi_{\ge E+\Delta E}^{(n)} \ket{\psi_{\le E}(t)}\|^2,
$
where $t=mT$ and $\ket{\psi_{\le E}(t)}=e^{-iH_\flo t}\ket{\psi_{\le E}}$.
Now, Theorem~\ref{FL:Metastability} is immediately followed by
\begin{align}
\| \Pi_{\ge E+\Delta E}^{(n_0)} \ket{\psi_{\le E}(t)}\|^2 \le & \left(6 t V_0  2^{-n_0}\right)^2. \label{energy_absorption_pre}
\end{align}
for $n=n_0$.
In this estimation, the probability does not depend on the amplitude of $\Delta E$; 
however, we expect that the probability of energy absorption should depend on the energy amplitude $\Delta E$.
Indeed, we can prove the following much stronger statement than \eqref{energy_absorption_pre}:

  \begin{figure}
\centering
\includegraphics[clip, scale=0.29]{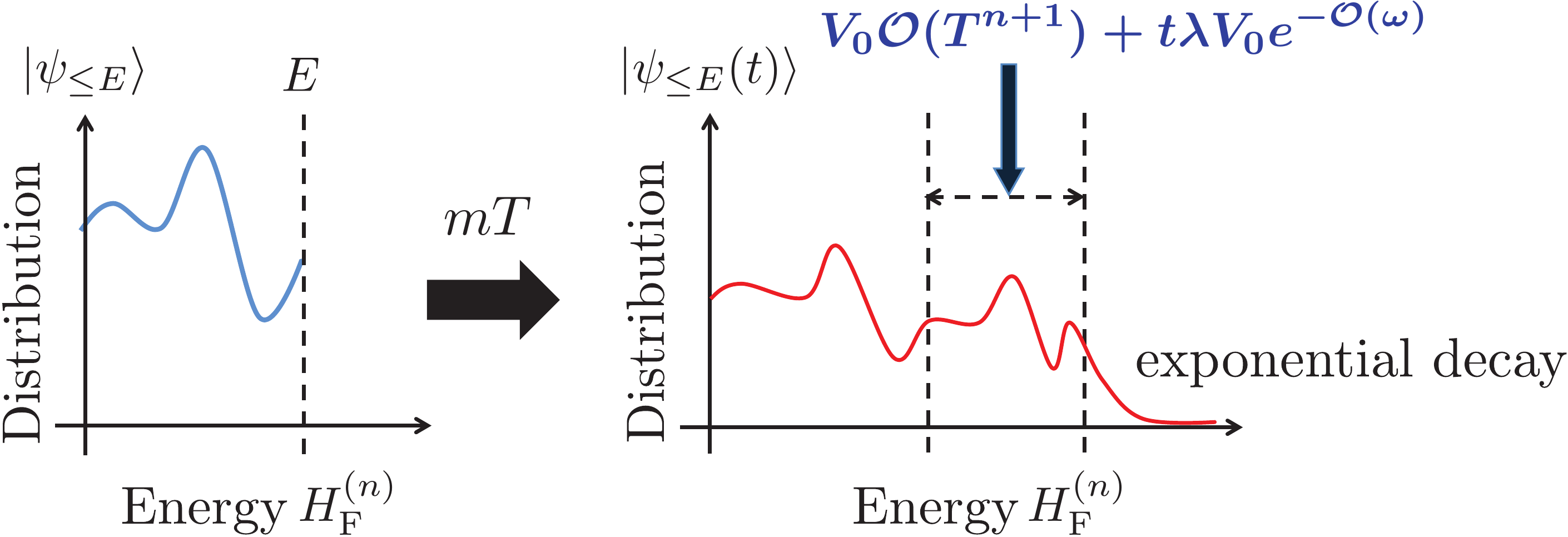}
\caption{Schematic picture of Theorem~\ref{FL:Heat absorption} for $H_\flo^{(n)}$.
Let us consider an initial state $\ket{\psi_{\le E}}$ which is in a superposition of energies below $E$ (blue curve). 
Then, after a time of $t=mT$, the energy distribution still localizes in the range $[0,E+V_0\orderof{T^{n+1}} + t \lambda V_0 e^{-\orderof{\omega}}]$, beyond which the distribution decays exponentially (red curve). 
}
\label{fig:FL_Energy_absorption}
\end{figure}

\begin{theorem}\label{FL:Heat absorption}
When we assume 
$
T \le \tau:= 1/(8\tilde{\lambda})
$
with $\tilde{\lambda}:=6k^2J$, 
the following inequality holds
\begin{align}
&\| \Pi_{\ge E+\Delta E}^{(n)} \ket{\psi_{\le E}(t)}\|^2 \notag \\
\le& \exp\left[2\tau \left( - \Delta E+ V_0 T^{n+1} \tilde{W}_{n+1}+22t  \lambda  V_0  2^{-n_0/2} \right)   \right]\notag 
\end{align}
with $t=mT$,
where $\tilde{W}_n$ is defined by
$
\tilde{W}_n:= 2 (4\lambda/3)^n  n! /(n+1)^2. 
$
We show the proof in Appendix~\ref{FL:proof_energy_localization}.
\end{theorem}
From Theorem~\ref{FL:Heat absorption}, we can identify the following two points (Fig.~\ref{fig:FL_Energy_absorption}):
\begin{enumerate}
\item{} After a time $t=mT$, the system can absorb energy at most 
$\Delta E \approx V_0 \orderof{T^{n+1}}+t \lambda V_0 e^{-\orderof{\omega}}$;   
that is, the energy absorption rate is exponentially suppressed for the frequency. 
\item{} The probability for large energy absorption decays exponentially or even faster as $\Delta E$ increases. 
\end{enumerate}
On the point~1., Ref.~\cite{abanin2015exponentially} has recently obtained the similar results in the linear response regime, which has been generalized in our recent paper~\cite{mori2015rigorous}.

\section{Relations to several phenomena} \label{FL:Relations to known results}

From our results, we can identify several properties which universally appear in periodically driven systems.
We here relate them to the existing results in literatures.

$~$

\noindent\textbf{1. Comparison with experimental time scale}.
Experimentally, the control parameter of the system is often given by $\hbar \omega/J$ (See references in~\cite{bukov2014universal}). 
Remember that the parameter $J$ characterizes the one-spin energy.
From Theorem~\ref{FL:Metastability}, the accuracy of the approximation is ensured as long as 
\begin{align}
m \lesssim 2^{\hbar/(16\lambda T)}=\exp \left( \frac{\log 2}{64}\cdot \frac{\hbar \omega}{J}\right ) \label{experimental_est}
\end{align}
with $k=2$ (i.e., two-body interaction), where we explicitly describe the Planck constant $\hbar$. 
Hence, the reliable time of the approximation sensitively depends on the coefficient of $\hbar \omega/J$, which is now given by $\log2/64$. 
When we consider an optical field as the periodic driving ($\omega\approx10^{12}$--$10^{16}$ Hz), our present estimation shows that, as long as $\hbar \omega/J \approx 10^3$--$10^4$, the lifetime $mT$ (or $m/\omega$) is comparable to the experimental time scale.
This condition still leaves room for improvement as we presently consider the most general setup in estimating the coefficient in \eqref{experimental_est}.

$~$

\noindent\textbf{2. Thermalization.} 
Based on our results, we give a general scenario for the thermalization process in periodically driven systems.
For arbitrary $n$ with $n\le n_0$, as shown in the inequality~\eqref{eq:general_n}, the time evolution of the $n$-th order truncated Floquet Hamiltonian approximates the exact time evolution up to $t\approx \orderof{\omega^{n+1}}$.
For larger $t$, the time evolution is not approximated by $H_\flo ^{(n)}$.
However,  up to $t\approx e^{\orderof{\omega}}$, $H_\flo ^{(n_0)}$ still gives a good approximation for the time evolution.
As long as $H_\flo ^{(n_0)}$ is ergodic, long-time unitary time evolution under $H_\flo ^{(n_0)}$ would result in an equilibrium state~\cite{PhysRevA.43.2046,PhysRevE.60.3949,PhysRevE.50.888,rigol2008thermalization}, that is,
\beq
\<\psi(t)|O|\psi(t)\>\simeq \tr ( O\rho_{\rm mc}^{(n_0)}),
\eeq
where $O$ is an arbitrary local operator and $\rho_{\rm mc}^{(n_0)}$ is the density matrix corresponding to the microcanonical ensemble of $H_\flo ^{(n_0)}$.
Because the difference between $H_\flo ^{(n)}$ and $H_\flo ^{(n_0)}$ is tiny for small $T$, the microcanonical ensemble of $H_\flo ^{(n)}$ would approximately equal to that of $H_\flo ^{(n_0)}$, $\rho_{\rm mc}^{(n)}\simeq\rho_{\rm mc}^{(n_0)}$ for any $n\leq n_0$.
We therefore expect that for sufficiently small $T$, we have
\beq
\<\psi(t)|O|\psi(t)\>\simeq \tr ( O\rho_{\rm mc}^{(n)}) \for  \forall n\le  n_0
\eeq
as long as $t\lesssim e^{\orderof{\omega}}$.
Although $\||\psi(t)\>-e^{-iH_\flo ^{(n)}t}|\psi\>\|$ is not small for $t\gg  \orderof{\omega^{n+1}}$, the system will reach a quasi-stationary state represented by the microcanonical ensemble of $H_\flo ^{(n)}$ and this quasi-stationary state persists at least up to $t\approx e^{\orderof{\omega}}$.
Thus, even the 0-th order truncated Floquet Hamiltonian $H_\flo ^{(0)}$ would well describe a quasi-stationary state of a periodically driven system up to an exponentially long time with respect to $\omega$.

This characterizes a kind of prethermalization in the sense that the system first relaxes to a quasi-thermal state $\rho_{\rm mc}^{(n_0)}$, which is far from the final equilibrium state, namely the infinite temperature state.
The mechanism of this prethermalization originated from the approximate freezing of the energy absorption and is qualitatively different from the 
established ones~\cite{
PhysRevLett.93.142002,
gring2012relaxation,
kaminishi2014entanglement,
bukov2015prethermal,
canovi2015stroboscopic}.

$~$

\noindent\textbf{3. Dissipative systems with periodic driving.} 
Our theorem is also applicable to an open system in which a small driven system is in contact with a large heat bath, as long as the total system is governed by a few-body Hamiltonian and the state vector of the total system obeys the Schr\"odinger equation.
According to the discussion in thermalization, the total system including the heat bath goes to the microcanonical ensemble, which implies that the system of interest reaches a quasi-stationary state described by the canonical ensemble of $H_\flo ^{(0)}$ (the Floquet-Gibbs state~\cite{Shirai2015,PhysRevE.79.051129}) with the inverse temperature $\beta$ identical to that of the thermal bath and this quasi-stationary state persists at least up to $t\approx e^{\orderof{\omega}}$.
At first glance, it seems contradictory to the fact that the long-time asymptotic state of the system of interest is in general \textit{not given} by the Floquet-Gibbs state even for large $\omega$~\cite{kohn2001periodic,Shirai2015,PhysRevE.61.4883,PhysRevE.89.012101,PhysRevE.90.012110} .
To resolve it, we should remember that the system of interest stays in the Floquet-Gibbs state \textit{for a fixed-time scale}, given by $e^{\orderof{\omega}}$.
 
Driven open systems are often treated by the Floquet-Born-Markov master equation~\cite{Shirai2015,PhysRevE.79.051129}, which is valid in the van Hove limit~\cite{van1955energy} (or the weak-coupling limit); as shown below, we cannot use the Floquet-Born-Markov master equation to grasp the finite-time behavior of the system.  
If the interaction Hamiltonian between the small system and the thermal bath is denoted by $\lambda H_I$, the limit of $\lambda\rightarrow 0$ and $t\rightarrow\infty$ with $\lambda^2t$ held fixed is called the van Hove limit.
Therefore, in the van Hove limit for a fixed value of $\omega$, the time scale of $t\approx e^{\orderof{\omega}}$ is always \textit{infinitesimally shorter} in comparison with the relaxation time, and thus the Floquet-Born-Markov master equation cannot capture the quasi-stability of the Floquet-Gibbs state, which has the lifetime of $e^{\orderof{\omega}}$.
This necessitates us to treat the dissipative system beyond the van-Hove limit in order to discuss the realistic relaxation process.
This necessitates us to treat the dissipative system beyond the van-Hove limit in order to discuss the realistic
relaxation process (see also the recent paper~\cite{shirai2015floquet}). 

%
%
%

$~$

\noindent\textbf{4. Numerical simulation in low-energy regime}.
Our analysis includes an implication on efficient numerical calculation to simulate periodically driven systems.
Our theorem implies the existence of approximate local conserved quantities, which is given by $H_\flo^{(n_0)}$.
Practically, we often cannot obtain the high-order Magnus terms for large systems, whereas only a few terms of the series give us a lot of useful information.
For example, let us choose the ground state of $H_\flo^{(0)}$ as the initial state. 
Then, after a time of $t\lesssim e^{\orderof{\omega}}$, Theorem~\ref{FL:Heat absorption} ensures that the energy is exponentially concentrated in the range of $[E_0, E_0+\orderof{T}V_0 )$, where we denote the ground state energy of $H_\flo^{(0)}$ by $E_0$.
This fact is beneficial in numerically simulating the low-energy periodic dynamics, because the entanglement area law is ensured for any low-energy state 
in various classes of Hamiltonians which are characterized by the heat capacity~\cite{brandao2014entanglement}; thereby, we can apply the matrix product state or the projected entangled pair state to simulate the dynamics up to the time $t\approx e^{\orderof{\omega}}$~\cite{perez2006matrix,verstraete2008matrix,PhysRevLett.91.147902,PhysRevLett.97.157202}.

$~$

\noindent\textbf{5. Dynamical localization.} 
When the dynamical localization occurs, the energy of the system cannot exceed a certain non-trivial energy bound.
From Theorem~\ref{FL:Heat absorption}, we can define $H_\flo^{(n_0)}$ as the approximately conserved energy for a long time of $t\approx e^{\orderof{\omega}}$; hence  
the dynamical localization generally occurs for large frequencies, but in the approximate sense.
We notice that the evidence of such behavior has been numerically reported in Ref.~\cite{d2013many}.

The dynamical localization often occurs in periodically driven systems, whereas the mechanism is not the same in all of the cases;
for example, the dynamical Anderson localization~\cite{
PhysRevLett.49.509}
 occurs because a periodic kick effectively causes random potential in momentum space, 
the dynamical freezing or coherent destruction of tunneling~\cite{
PhysRevLett.67.516,
mondal2012dynamics,
PhysRevB.82.172402,
PhysRevB.90.174407},
originate from the resonant interference effect at particular frequencies, the many-body localization~\cite{
PhysRevLett.111.127201,
PhysRevB.90.174202,
PhysRevB.91.085425,
PonteMBL2015,
MBL_Analabha_2015,
abanin2014theory}
 is related to the local independence of spins which comes from disorder in the Hamiltonian, etc. 

In the present case, we can provide a qualitative explanation 
in terms of the following viewpoint. 
In the periodically driven system, the unit of energy which the system can absorb is roughly quantized  by $\omega$, while 
the energy associated to one spin is bounded by $J$ as in \eqref{eq:extensiveness}.
Thus, for the energy absorption of $\orderof{\omega}$, many spins (i.e., $\omega/J$ spins) should work synergistically in one period. 
However, such many-body quantum effect is exponentially suppressed in a short-time evolution for the few-body Hamiltonians~\cite{kuwahara2015locality}.
In this way, the energy absorption rate is bounded from above by $e^{-\orderof{\omega}}$; this mechanism of the (quasi-) dynamical localization is distinct from the existing ones.

$~$

%
%

\noindent\textbf{6. Small integrability breaking}. 
We here consider the case where a small integrability breaking term $\epsilon V$ is added to an integrable Hamiltonian $H^\ast(t)=H^\ast(t+T)$, that is, $H(t)=H^\ast(t)+\epsilon V$, where $\epsilon V$ is time-independent.
Then, the Floquet Hamiltonian $H_\flo$ is formally expanded as 
\begin{align}
H_\flo :=H_\flo^\ast + \epsilon H_{1,\flo} + \epsilon^2 H_{2,\flo}  + \cdots , \notag 
\end{align}
where $H_\flo^\ast $ is the Floquet Hamiltonian with respect to $H^\ast(t)$, which might have a non-trivial form far from the random matrix. For example, let us choose bilinear fermionic Hamiltonians (e.g., $XY$ model) as the unperturbed Hamiltonian $H^\ast(t)$. 
Then, the Floquet Hamiltonian $H_\flo^\ast $ is always far from the random matrix~\cite{PhysRevLett.112.150401,PhysRevLett.109.257201}; that is, heat up to the infinite temperature never occurs.
On the other hand, an infinitesimal perturbation  $\epsilon V$ can cause the heating after sufficiently long time, which  
kills all the specific properties which are originated from $H_\flo^\ast$. 
We discuss the time-scale for which the integrability breaking term becomes dominant.

In order to give the effect of $\epsilon V$, we further expand $H_{j,\flo}$ with respect to $T$ as 
$ \epsilon^j H_{j,\flo}^{(n_0)}=  \epsilon^j \sum_{m=0}^{n_0} T^m\Omega_{j,m}$, where $\{\Omega_{j,m}\}$ can be determined by the use of the FM expansion. 
Then, our result ensures that the system is governed by the following effective Hamiltonian 
\begin{align}
H_\flo^{(n_0)} =H_\flo^\ast +\epsilon  \sum_{m=0}^{n_0} T^m ( \Omega_{1,m} + \epsilon \Omega_{2,m}  + \cdots ) \notag 
\end{align}
as long as $t\lesssim e^{\orderof{\omega}}/\epsilon$.
This implies the following: 
\begin{enumerate}
\renewcommand{\labelenumi}{\roman{enumi})}
\item{}  Up to the time $t\approx 1/\epsilon$, the system is well approximated by the integrable Hamiltonian $H_\flo^\ast$ and described by the generalized Gibbs ensemble~\cite{PhysRevLett.112.150401}. 
\item{} The non-integrable terms $\sum_{m=0}^{n_0} T^m ( \Omega_{1,m} + \epsilon \Omega_{2,m}+\cdots)$ influence the dynamics in the time scale of $1/\epsilon \ll t \lesssim e^{\orderof{\omega}}/\epsilon$ and reduce the system to the Gibbs ensemble by $H_\flo^{(n_0)}$. 
In this time scale, the effective Hamiltonian $H_\flo^{(n_0)}$ is still close to $H_\flo^\ast$, and hence
the characteristic properties of the system may strongly depends on $H_\flo^\ast$.
\item{} The divergence terms of the FM expansion finally kill all the non-trivial structure of $H_\flo^\ast $ for $t \gg e^{\orderof{\omega}}/\epsilon$ and heat up the system to the infinite temperature state. 
\end{enumerate}

\section{Summary and future perspective} \label{FL:Summary}

In summary, we have worked on the generic transient dynamical behavior of periodically driven systems.
In the thermodynamic limit, the Floquet-Magnus expansion~\eqref{eq:FM_finite} usually diverges for $n\to \infty$, whereas it 
converges up to a certain order of $\orderof{\omega}$ as shown in Lemma~\ref{FL:FM_bound}. 
Our main results characterize the time-scale for which the truncation of the FM expansion provides a good approximation: $e^{-iH_\flo T} \simeq e^{-iH_\flo^{(n_0)}T}$ with an exponentially small error for the frequency.
It is worth noting that our results can be applied to macroscopic systems ($N\to \infty)$ and 
are highly universal in that we consider generic few-body Hamiltonians without specific assumptions.
Our results have given the first theoretical step to understand the transient dynamics systematically.

On the other hand, our work has left several open problems.
First, the main theorems only refer to the upper bound on the error of approximation, and hence the infinite-time behavior is still an open problem. 
From the convergence condition of the FM expansion, we can ensure that only when $\omega \gtrsim N$ the exact Floquet Hamiltonian is far from the random matrix. 
On the other hand, our results indicate that for $\omega \gtrsim \log N$ the truncated FM expansion $H_\flo^{(n_0)}$ gives a very good approximation 
at least for the unitary operator, i.e., $e^{-iH_\flo T} \simeq e^{-iH_\flo^{(n_0)} T} $.
Hence, we expect that we might obtain a looser condition than $\omega \gtrsim N$ for the exact Floquet Hamiltonian to have a well-defined form. 
Indeed, the recent results support this expectation; for example, the finding by D'Alessio and Rigol~\cite{Rigol_DAlessio2014} implies
that the long-time behavior of a periodically driven lattice system qualitatively changes at around $\omega \gtrsim \sqrt{N}$.

Second, it would be an interesting direction to extend our present theory to Liouvillian dynamics in dissipative driven quantum systems. 
The dissipation causes non-trivial quantum phenomena which cannot be simply explained and has recently attract much attention 
not only in fundamental~\cite{PhysRevA.91.051601,PhysRevLett.105.227001,PhysRevLett.95.073003} but also in experimental aspects~\cite{PhysRevX.4.031043,diehl2011topology,
PhysRevLett.114.056801}. 
We believe that our framework will be applicable to generic Liouvillian dynamics with time periodicity; 
here, we can define an effective Liouvillian operator. 
The problem is not straightforward because the basic assumptions of \eqref{few_body_Hamiltonian_eq}
and \eqref{eq:extensiveness} are not trivially ensured for generic Liouvillian operators. 
Even though, several techniques applied here, such as the Lieb-Robinson bound, can be also applicable to various classes of dissipative dynamics~\cite{Cubbit_dissipative,PhysRevLett.104.190401}.  
We expect that the qualitative properties are the same between the effective Liouvillian operator and the Floquet
Hamiltonian in closed systems.

Third, related to the discussions~2 and 3 in Sec.~\ref{FL:Relations to known results}, 
it is an intriguing problem whether our scenario on the thermalization can be numerically (or experimentally) observed or not.
Without a heat bath, a periodically driven system is conjectured to show two-step relaxation process; it first goes to a prethermalized state which is characterized by the effective Hamiltonian $H_\flo^{(n_0)}$ and then relaxes to the final steady state, i.e. the infinite temperature state.
On the other hand, in the presence of a heat bath, the quasi-steady state characterized by $H_\flo^{(n_0)}$ might have infinite lifetime, where  
the energy emission to the heat bath should be comparable to the energy absorption from the external driving.
Although  we will have to treat a considerable number of spins in the numerical demonstration, 
we expect that we can simulate the relaxation processes efficiently in the low-energy regime according 
to the discussion~4 in Sec.~\ref{FL:Relations to known results}.


\section{Methods} \label{FL:Proof of our results}
In this section, we show the proofs of our main theorems.
We first show a decomposition of the Floquet operator as an important stepping stone to the proof.
Based on it, we give outlines of the proofs and defer the details of the calculations to Appendix sections.

\subsection{Decomposition of the Floquet operator} \label{decomp_FL_op}
As a main idea for the proofs, we introduce a decomposition of the total unitary operator:
\begin{align}
U_\flo = e^{-iH_0T} U_N U_{N-1}  \cdots  U_2 U_1  \label{unitary_decomp_total}
\end{align}
and 
\begin{align}
U^{(n_0)}_\flo = e^{-iH_0T} U^{(n_0)}_N U^{(n_0)}_{N-1} \cdots  U^{(n_0)}_2 U^{(n_0)}_1 , \label{unitary_decomp_total_effective}
\end{align}
where we denote the unitary operators $e^{-iH_\flo T}$ and $e^{-iH_\flo^{(n)}T}$ by $U_\flo$ and $U_\flo^{(n)}$, respectively.
In this decomposition, we first separate the driving Hamiltonian $V(t)$ into $N$ pieces, namely $\{V_i(t)\}_{i=1}^N$, and aim to 
treat their driving effects separately. 
As shown below, the decomposed unitary operators $\{U_i\}_{i=1}^N$ correspond to the contributions by $\{V_i(t)\}_{i=1}^N$, respectively.

We first introduce the notations of $\Lambda_{0}:=\emptyset$, $\Lambda_{i}:=\{1,2,\ldots,i\}$ and
\begin{align}
&V_i(t) := \sum_{X:X \ni i, X\cap \Lambda_{i-1}=\emptyset } v_X(t) ,     \notag  \\
&\tilde{H}_i(t) := H_0+ \sum_{j= i+1}^NV_j(t)  \label{Definition_of_bar_H_m_t}
\end{align}
for $i=0,1,\ldots,N$, where 
$\tilde{H}_{0}(t)=H(t)$, $\tilde{H}_{N}(t)=H_0$ and $\tilde{H}_i(t)=V_{i+1}(t)+\tilde{H}_{i+1}(t)$. 
Note that $V_i(t)$ is the driving operator which contains the spin $i$ but not contains the spins $\{1,2,\cdots,i-1\}$.
Using the above notations, we first decompose the total Hamiltonian into $H(t)=V_1(t)+\tilde{H}_1(t)$, which reduces
 the total unitary operator to
$
 U_\flo  =U_{\flo,1}(T) U_1  \notag 
$
with
$U_{\flo,1}(t):= \mathcal{T} \bigl[ e^{-i\int_0^t \tilde{H}_1(t) dt} \bigr]$ and 
$
U_1= \mathcal{T} \bigl[ e^{-i\int_0^T U_{\flo,1}^\dagger(t)  V_1(t) U_{\flo,1}(t) dt} \bigr].
$
Here, $U_{\flo,1}(T) $ is the Floquet operator with respect to the Hamiltonian $\tilde{H}_1(t)$.
We can then decompose as $\tilde{H}_1(t)=V_2(t)+\tilde{H}_2(t)$ and define $U_{\flo,1}(T)=U_{\flo,2}(T) U_2$ in the same way.
By repeating this process, we decompose the total unitary operator $U_\flo$ as in Eq.~\eqref{unitary_decomp_total}, 
where the decomposed unitary operators are given by $U_{\flo,i-1}(T)=U_{\flo,i}(T) U_{i}$ such that 
\begin{align}
&U_i:= \mathcal{T} \bigl[ e^{-i\int_0^T U^\dagger_{\flo,i}(t) V_i(t ) U_{\flo,i}(t) dt} \bigr] ,\notag \\
&U_{\flo,i}(t) := \mathcal{T} \bigl[ e^{-i\int_0^t \tilde{H}_{i}(t) dt} \bigr] =e^{-iH_0T}U_N  \cdots   U_{i+1}  \label{unitary_decomp_total_detail}
\end{align}
for $i=1,2,\ldots,N$.

Second, we define $\tilde{H}^{(n_0)}_i$ as the truncated FM expansion at $n_0$th order with respect to the time-dependent Hamiltonian $\tilde{H}_i(t)$, 
where $\tilde{H}^{(n_0)}_0=H_\flo^{(n_0)}$ and $\tilde{H}^{(n_0)}_N=H_0$ because of $\tilde{H}_0(t)=H(t)$ and $\tilde{H}_N(t)=H_0$, respectively.
We then introduce the notation of $V_i^{(n_0)}:=\tilde{H}^{(n_0)}_{i-1}-\tilde{H}^{(n_0)}_i$ and have 
\begin{align}
\tilde{H}^{(n_0)}_i = H_0+\sum_{j=i+1}^N V_j^{(n_0)}. \label{Definition_of_bar_H_m_t_n0}
\end{align}
Based on the above notations, we also decompose the effective unitary operator $U^{(n_0)}_\flo$ in the similar way to \eqref{unitary_decomp_total_detail}:
\begin{align}
&U_i^{(n_0)}:= \mathcal{T} \bigl[ e^{-i\int_0^T U_{\flo,i}^{(n_0)\dagger}(t)V_{i}^{(n_0)}U_{\flo,i}^{(n_0)}(t) dt} \bigr] ,\notag \\ 
&U_{\flo,i}^{(n_0)}(t) := e^{-i\tilde{H}_i^{(n_0)}t}=e^{-iH_0T}U_N^{(n_0)}  \cdots   U^{(n_0)}_{i+1}  \label{unitary_decomp_total_effective_detail}
\end{align}
for $i=1,2,\ldots,N$.  Note that the definition~\eqref{unitary_decomp_total_effective_detail} implies $U_{\flo,i-1}^{(n_0)}(T)=U_{\flo,i}^{(n_0)}(T)U_{i}^{(n_0)}$.

For the convenience, we often include $H_0$ into $V(t)$; that is, we denote $H(t)=H'_0 + V'(t)$ with $H'_0=\hat{0}$ and $V'(t)=H(t)$.
In this case, the decomposition \eqref{unitary_decomp_total} reads
\begin{align}
U_\flo = U'_NU'_{N-1}  \cdots U'_2  U'_1 . \label{unitary_decomp_total2}
\end{align}
Then, the definition~\eqref{Definition_of_bar_H_m_t} implies that the Hamiltonian $\tilde{H}'_i(t) =\sum_{j\ge i+1} V'_j(t)$ contains the spins $\{i+1,i+2\ldots, N\}$, and hence 
from \eqref{unitary_decomp_total_detail}, the unitary $U'_i$ is supported in the spin set $\{i,i+1\ldots, N\}$.


\subsection{Basic lemma on the decomposed unitary operators}

For each of the decomposed Floquet operators $\{U_i\}_{i=1}^N$, we can derive the following lemma;
we utilize it as a central technique for the subsequent proofs.

\begin{lemma}\label{FL:Metastability_local}
Let $\{U_i\}_{i=1}^N$ and $\{U_i^{(n_0)}\}_{i=1}^N$ be the decomposed unitary operators given in 
\eqref{unitary_decomp_total_detail} and \eqref{unitary_decomp_total_effective_detail}. 
Then, each of the unitary operators $\{U_i\}_{i=1}^N$ is close to $\{U_i^{(n_0)}\}_{i=1}^N$ with $n_0=\left \lfloor\frac{1}{16\lambda T} \right\rfloor$ as follows:
\begin{align}
\| U_i-  U_i^{(n_0)} \| \le 6\Vn_i T 2^{-n_0},
\label{eq:theorem_local}
\end{align}
where $\lambda$ is defined in Eq.~\eqref{Notations_V_0_lambda} and we define $\{\Vn_i\}_{i=1}^N$ as
\begin{align}
\Vn_i:=\sum_{X:X \ni i, X\cap \Lambda_{i-1}=\emptyset }\frac{1}{T}\int_0^T  \| v_X(t) \| \le J .\label{Def_Vni}
\end{align}
Notice that because of Eq.~\eqref{Notations_V_0_lambda}
\begin{align}
\sum_{i=1}^N  \Vn_i = \sum_{X\in \Lambda}\frac{1}{T}\int_0^T  \| v_X(t) \|  =  V_0. \label{summation_Vn_i}
\end{align} 
\end{lemma}

{~}\\
\textit{Proof of Lemma~\ref{FL:Metastability_local}}.
For the proof, we start from the equalities $U_{\flo,i-1}(T)=U_{\flo,i}(T) U_i$ and $U_{\flo,i-1}^{(n_0)}(T)=U_{\flo,i}^{(n_0)}(T) U_i^{(n_0)}$, from which 
we expand $U_i$ and $U^{(n_0)}_i$ with respect to the period $T$:
\begin{align}
&U_i= U_{\flo,i}^\dagger(T)  U_{\flo,i-1}(T) = \sum_{n=0}^{\infty} T^n \Phi_{i,n}, \notag \\
&U^{(n_0)}_i =  U_{\flo,i}^{(n_0)\dagger}(T) U_{\flo,i-1}^{(n_0)}(T)  = \sum_{n=0}^{\infty} T^n \Phi^{(n_0)}_{i,n} , \label{expansion_U_1_U_1_n0}
\end{align}
where the terms $\Phi_{1,n} $ and $\Phi^{(n_0)}_{1,n}$ are invariant for the scaling $T\to sT$, $H(t)\to H(t/s)$ with $s$ a positive constant.

Now, the operator $U_{\flo,i}^{(n_0)}(T)$ comes from the FM expansion for the Hamiltonian $\tilde{H}_{i}(t)$. 
Hence, from the basic property of the FM expansion, 
when we expand $U_{\flo,i}(T)$ and $U_{\flo,i}^{(n_0)}(T)$ with respect to $T$ as 
$
U_{\flo,i}(T)  = \sum_{n=0}^{\infty} T^n \mathcal{U}_{i,n}$ and $U_{\flo,i}^{(n_0)}(T)   =\sum_{n=0}^{\infty} T^n \mathcal{U}^{(n_0)}_{i,n}
$, respectively, 
the expanded terms are identical up to the order $n_0$, namely 
$
\mathcal{U}_{i,n} = \mathcal{U}^{(n_0)}_{i,n} 
$
for $n\le n_0$.
By applying the above equality to \eqref{expansion_U_1_U_1_n0}, we also obtain
 \begin{align}
\Phi_{i,n} = \Phi^{(n_0)}_{i,n} \for n\le n_0,\notag 
\end{align}
which yields
\begin{align}
\| U_i-  U^{(n_0)}_i \| \le   \sum_{n=n_0+1}^{\infty} T^n (\| \Phi_{i,n}\| +\| \Phi^{(n_0)}_{i,n}\|). \label{proof_basic_1}
\end{align}

We then evaluate the upper bound of $\| \Phi_{i,n}\|$ and $\|\Phi^{(n_0)}_{i,n}\|$ based on the explicit 
forms of \eqref{unitary_decomp_total_detail} and \eqref{unitary_decomp_total_effective_detail} for $U_i$ and $U^{(n_0)}_i $, respectively.    
After straightforward but rather technical calculations, we can obtain the following inequalities for the norms of expanded terms:
\begin{align}
 &T^n \| \Phi_{i,n} (T)\| \le \Vn_i T  e^{1/ (2k)}  2^{-n+1}    , \label{proof_basic_Phi_n}\\
 &T^n \| \Phi^{(n_0)}_{i,n} (T)\|\le  \Vn_i T e^{1/(8k) } 2^{-n+1} .\label{bound_for_T_n_Phi_n_0} 
\end{align}
We show the derivations in Appendix~\ref{Sec:tilde_Gamma_l_S}.

By applying the inequalities~\eqref{proof_basic_Phi_n} and \eqref{bound_for_T_n_Phi_n_0} to \eqref{proof_basic_1}, we obtain
 \begin{align}
\| U_i-  U^{(n_0)}_i \| &\le   2\Vn_i T  2^{-n_0} (e^{1/2}   + e^{1/8 })  \le 6\Vn_i T 2^{-n_0},\notag 
\end{align}
where we use $k\ge1$ from the definition of \eqref{few_body_Hamiltonian_eq}.
This completes the proof.
$\blacksquare$

\subsection{Proof of Theorem~\ref{FL:Metastability}}\label{FL:Proof of our results_thm1}

By applying Lemma~\ref{FL:Metastability_local} to the decompositions~\eqref{unitary_decomp_total} and \eqref{unitary_decomp_total2}, 
the proof is immediately followed by 
\begin{align}
&\| U_\flo-  U_\flo^{(n_0)} \| \le \|U_1U_2\cdots U_N-   U_1^{(n_0)}U_2^{(n_0)} \cdots U_N^{(n_0)}\|  \notag \\
&\le  \|U_1-U_1^{(n_0)} \|+   \| U_2\cdots U_N - U_2^{(n_0)} \cdots U_N^{(n_0)} \| \notag \\
&\le \sum_{i=1}^N  \|U_i-U_i^{(n_0)} \| \le  6T 2^{-n_0} \sum_{i=1}^N \Vn_i  .  \label{Theorem_1_first_ineq}
\end{align}
By using Eq.~\eqref{summation_Vn_i} in the above inequality, we complete the proof.
$\blacksquare$

\subsection{Proof of Theorem~\ref{FL:Metastability4}} \label{FL:Proof of our results_thm2}

For the proof, we first show the following corollary on the time evolution of a local operator:
\begin{corollary}\label{FL:Metastability3}
Let $O_L$ be an arbitrary operator supported in a region $L$.
Then, the exact time evolution of $O_L(T)$ by $U_\flo$ is close to the approximate time evolution $O_L^{(n_0)}(T)$ by $U_\flo^{(n_0)}$: 
\begin{align}
\| O_L(T)- O_L^{(n_0)}(T) \| \le 12J|L| T  2^{-n_0}\|O_L\|. \notag 
\end{align}
\end{corollary}

~\\
\textit{Proof of corollary~\ref{FL:Metastability3}}.
In order to prove the corollary, we consider the decomposition~\eqref{unitary_decomp_total2}, and then 
we can obtain the same inequality as \eqref{eq:theorem_local}:
\begin{align}
\| U'_i-  U_i^{'(n_0)} \| \le   6J T 2^{-n_0}. \label{eq:theorem_local__second}
\end{align}
We now label spins in the region $L$ by $\{1,2,\ldots, |L|\}$. 
Remembering that from the definition~\eqref{unitary_decomp_total2} $\{U'_i\}_{i=|L|+1}^N$ are supported in the region $L^\co$, 
we have $O_L(T)=U_L^{'\dagger} O_L U'_L $ with the notations of $U'_L =U'_{|L|}  \cdots U_2'U_1' $; 
in the same way, we also have $O_L^{(n_0)}(T) =U_L^{'(n_0)\dagger} O_L  U_L^{'(n_0)}$.

From the inequality~\eqref{eq:theorem_local__second}, we obtain
$
\| U'_{L} - U_L^{'(n_0)}\| \le 6J |L| T 2^{-n_0}
$
in the similar way to the inequality~\eqref{Theorem_1_first_ineq}.
We thus obtain
\begin{align}
&\| O_L(T)- O_L^{(n_0)}(T) \|=\| U_L^{'\dagger} O_L U'_L -  U_L^{'(n_0)\dagger} O_L U_L^{'(n_0)} \| \notag \\
&\le2 \| U'_L - U_L^{'(n_0)}\|  \cdot \|O_L \| \le 12J |L| T 2^{-n_0}  \|O_L\|.\notag 
\end{align}
This completes the proof of Corollary~\ref{FL:Metastability3}. $\blacksquare$

~\\
\textit{Proof of Theorem~\ref{FL:Metastability4}}.
From the definition~\eqref{def_rho_mT_2}, $\rho_L(t) -\rho_L^{(n_0)}(t)$ is an Hermitian operator, and hence we can always find an operator $O_L$ with $\|O_L\|=1$ such that
 \begin{align}
&\| \rho_L(t) -\rho_L^{(n_0)}(t)\|_1=\tr \left[O_L  \left(\rho_L(t ) -\rho_L^{(n_0)}(t)\right)\right] \notag \\
&= \tr \left[\rho  \left(O_L(t) -O_L^{(n_0)}(t) \right)\right]  \le \| O_L(t) -O_L^{(n_0)}(t)\| , \notag 
\end{align}
where we use $|\tr (BA)| \le \|B\|_1 \cdot \|A\|$ for arbitrary operators $A$ and $B$; note that $\|\rho\|_1=1$.
We therefore have to calculate the upper bound of
$
\| O_L(mT) -O_L^{(n_0)}(mT)\|.
$

 \begin{figure}
\centering

\includegraphics[clip, scale=0.3]{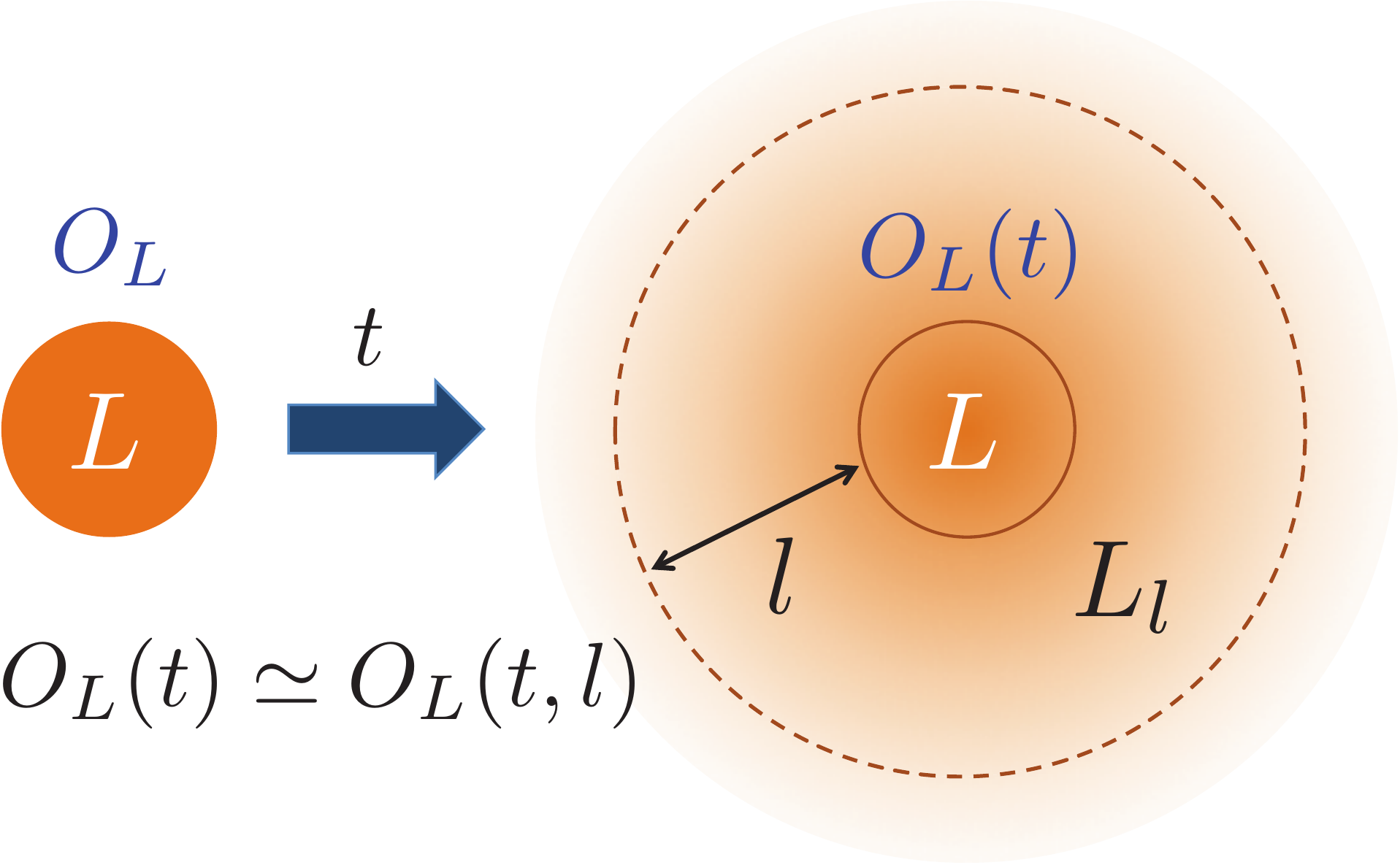}

\caption{Schematic picture of the Lieb-Robinson bound. 
We consider an operator $O_L$ which is initially supported in a region $L$. 
after a short time, the operator $O_L(t)$ is no longer supported in the region $L$. 
However, we expect that the operator $O_L(t)$ still remains in a region closed to $L$; that is, it is approximated in some region $L_l$ having distance $l$ from $L$.
The Lieb-Robinson bound gives the accuracy of this approximation.
}
\label{fig:LR_bound}
\end{figure}

For the evaluation of the norm, we can apply Corollary~\ref{FL:Metastability3} to $\| O_L(T) -O_L^{(n_0)}(T)\|$, whereas for $\| O_L(2T) -O_L^{(n_0)}(2T)\|=\| U_\flo O_L(T)U_\flo^\dagger   -U_\flo^{(n_0)}  O_L^{(n_0)}(T)U_\flo^{(n_0)\dagger}\|$, the corollary cannot be applied directly; that is,
the operator $O_L(T)$ is \textit{no longer supported in the region $L$} due to the time evolution (Fig.~\ref{fig:LR_bound}). 
We thus would like to know how accurate the operator $O_L(t)$ can be approximated by an operator acting on spins which is close to $L$. 
For this purpose, we denote the set of spins having distance at most $l$ from $L$ by $L_l$, namely  $\dist (L,i) \le l$ for $\forall i \in L_l $. 
We define the following operator $O_L(t,l)$ in the region $L_l$ which approximates the operator $O_L(t)$:
 \begin{align}
O_L(t,l) := \frac{1}{\tr  (\hat{1}_{L_l^\co})}\tr_{L_l^\co} [ O_L(t)]\otimes \hat{1}_{L_l^\co}.\notag 
\end{align}
From Ref.~\cite{PhysRevLett.97.050401}, the Lieb-Robinson bound gives the error of the approximation as
 \begin{align}
\|O_L(t)-  O_L(t,l)\|\le |L| \LG(l,t), \label{LR_bound_region}
\end{align}
where $\LG(l,t)$ has been defined in Eq.~\eqref{Lieb_Robinson_bound_G}.

We now define $s_m := \| O_L(mT) -O_L^{(n_0)}(mT)\|$
 and  obtain for $s_{m+1}$
 \begin{align}
s_{m+1}
\le& \| U_\flo^\dagger   O_L(mT) U_\flo - U_\flo^{(n_0)\dagger}   O_L(mT)  U_\flo^{(n_0)} \|   \notag \\
&+ \| O_L(mT) -O_L^{(n_0)}(mT)\| \notag \\
\le& \| U_\flo^\dagger  O_L(mT,l ) U_\flo  - U_\flo^{(n_0)\dagger}   O_L(mT,l)  U_\flo^{(n_0)} \|  \notag \\
& + 2\|O_L(mT)-  O_L(mT ,l)\| + s_m\notag \\
\le& 12J |L_l| T  2^{-n_0}  +  2 |L| \LG(l,mT) +s_m, \notag 
 \end{align}
 where we apply Corollary~\ref{FL:Metastability3} and \eqref{LR_bound_region}  in the last inequality.
We thus obtain 
 \begin{align}
s_{m} &\le 12J |L_l| T m 2^{-n_0}   +  2 |L| m \LG(l,mT) , \label{FL_S_m_ineq}
 \end{align}
where we use $s_0 = 0$ and $\LG(l,mT) \le  \LG(l,m'T) $ for $m'\ge m$. 
Let us choose $l$ such that $ |L_l| =2^{n_0/2} |L|$; in this case, because we consider $D$-dimensional systems, we have $l=\const \cdot 2^{n_0/(2D)}= e^{\orderof{\omega/D}}$.
Then, the inequality~\eqref{FL_S_m_ineq} reduces to \eqref {eq:theorem4}.
We thus prove Theorem~\ref{FL:Metastability4}.
$\blacksquare$

\section{Acknowledgement}

We thank Yuji Shirai and Yoshihito Hotta for valuable comments on the manuscript.
TK acknowledges the support from JSPS grant no.~2611111. 
TM was supported by the JSPS Core-to-Core
Program ``Non-equilibrium dynamics of soft matter and
information'' and JSPS KAKENHI Grant No.~15K17718.
K.S. was supported by MEXT grant no.~ 25103003.

\textit{Note added.} In updating the previous version (arXiv:1508.05797) to this version,
related results  in a different approach have appeared~\cite{abanin2015asymptotic,abanin2015asymptotic2}.

\appendix
\section{Proof of Corollary~\ref{FL:m_period_theorem}} \label{Proof of Corollary1}

The inequality~(\ref{eq:general_n}) is shown by dividing as
\begin{align}
\| e^{-iH_\flo  T} -e^{-iH_\flo ^{(n)}T} \| \leq & \| e^{-iH_\flo  T}-e^{-iH_\flo ^{(n_0)}T} \| \notag \\
+&\| e^{-iH_\flo^{(n_0)}T}-e^{-iH_\flo^{(n)}T} \| .
\label{eq:general_n_proof1}
\end{align}
The first term is bounded by \eqref{eq:theorem}.
From Eq.~(\ref{eq:norm_Omega}), the second term is bounded from above by
$
\|H_\flo ^{(n_0)}-H_\flo ^{(n)}\|T \leq\sum_{m=n+1}^{n_0}\|\Omega_m\|T^{m+1}
 \leq\sum_{m=n+1}^{n_0} \frac{V_0 T}{(m+1)^2} (\lambda T)^m m!.
$
For $n+1\le m\le n_0$, we have $m!\leq n_0^{m-n-1}(n+1)!$; hence, we obtain
\begin{align}
&\|e^{-iH_\flo ^{(n_0)}T}-e^{-iH_\flo ^{(n)}T}\|
\le \overline{\Omega}_{n+1}T^{n+2},
\label{eq:general_n_proof2}
\end{align}
where we use $1/(m+1)^2\le 1/(n+2)^2$ for $m\ge n+1$ and $\lambda T n_0 \le 1/16$.
By combining the inequality~\eqref{eq:general_n_proof1} with \eqref{eq:theorem} and \eqref{eq:general_n_proof2}, we prove Corollary~\ref{FL:m_period_theorem}.
$\blacksquare$

\section{Basic properties of multi-commutator}

For the analysis of few-body observables, we often treat the multi-commutators.
Here, we show three lemmas as useful technical tools in the analysis.  
For the convenience, we introduce two parameters $(k_A,J_A)$ to characterize basic properties of a generic operator $A$.
We, in the following, refer to that an operator $A$ is $k_A$-local and $J_A$-extensive when
\begin{align}
&A= \sum_{|X|\le k_A} a_X \quad \textrm{($k_A$-local)}, \notag \\
&\sum_{X:X\ni i} \|a_X\| \le J_A \quad \textrm{($J_A$-extensive)}, 
\label{k_local_extensiveness}
\end{align}
for $\forall i\in \Lambda$. 
For example, the Hamiltonian given by \eqref{eq:extensiveness} with \eqref{few_body_Hamiltonian_eq} is $k$-local and $J$-extensive.

For the norm of multi-commutators, we can prove the following lemma:
\begin{lemma} \label{norm_multi_commutator1}
Let $\{A_i\}_{i=1}^n$ be $k_i$-local and $J_{i}$-extensive, respectively, and 
$O_L$ be an arbitrary operator supported in a region $L$.
Then, the norm of the multi-commutator 
\begin{align}
\{A_n,A_{n-1},\ldots , A_1,O_L\}:= [A_n,[A_{n-1}, [\cdots, [A_1,O_L ]] \notag
\end{align}
is bounded from above by
\begin{align}
\| \{A_n,A_{n-1},\ldots , A_1,O_L \}\| \le \prod_{m=1}^n (2  J_{m} K_m) \|O_L\|, \label{fundamental_ineq}
\end{align}
where $K_m := |L|+ \sum_{i\le m-1} k_i $.
\end{lemma}

{~}\\
\textit{Proof of Lemma~\ref{norm_multi_commutator1}}.
We start from the following expression for the multi-commutator~\cite{arad2014connecting}:
\begin{align}
&\{A_n,A_{n-1},\ldots , A_1,O_L \} \notag \\
=& \sum_{X_1|L}\ \ \sum_{X_2|(L,X_1)} \ \ \sum_{X_3|(L,X_1,X_2 )} \cdots
     \sum_{X_n|(L, X_1 X_2 \ldots,X_{n-1})} \notag \\
     &\{a^{(n)}_{X_n}, a^{(n-1)}_{X_{n-1}}, \ldots,   a^{(1)}_{X_1},O_L \},\notag 
\end{align}
where we define 
$
A_m :=  \sum_{|X| \le k_m } a^{(m)}_{X}
$
for $m=1,2,\ldots,n$ and $X| (L, X_1,X_2, \ldots,X_m)$ denotes the set of $\{ X: X\cap ( L \cup X_1 \cup X_2 \cup \cdots \cup X_m ) \neq \emptyset\}$.
We thus obtain
\begin{align}
&\| \{A_n,A_{n-1},\ldots , A_1,O_L \}\| \notag \\
\le&  2^n \sum_{X_1|L} \cdots \sum_{X_n|(L, X_1 X_2 \ldots,X_{n-1})} 
\| a^{(n)}_{X_n}\|   \cdots \| a^{(1)}_{X_1}\| \cdot \|O_L\|.\notag
\end{align}
For $m=1,2,3,\ldots, n$, we have
\begin{align}
&\sum_{X_m|(L, X_1 X_2 \ldots,X_{m-1})} \| a^{(m)}_{X_m}\| \notag \\
\le& \sum_{i : i \in( L\cup X_1 \cup X_2 \cup \cdots \cup X_{m-1} ) }  \sum_{X_m : X_m \ni i}\| a^{(m)}_{X_m}\| \notag \\
\le&  \sum_{i : i \in( L\cup X_1 \cup X_2 \cup \cdots \cup X_{m-1} ) }   J_{m}  \notag \\
\le& J_{m} \biggl ( |L|+ \sum_{i\le m-1} k_i \biggr) = K_m J_{m} ,\notag
\end{align}
where $J_m$-extensiveness~\eqref{k_local_extensiveness} implies the inequality $\sum_{X_m : X_m \ni i}\| a^{(m)}_{X_m}\|  \le J_m$.
We thus prove the lemma. $ \blacksquare$

As a related lemma, we also utilize the following lemma (See Ref~\cite{kuwahara2015locality} for the proof): 
\begin{lemma} \label{norm_lemma_commutator2}
Let $H$ be a $k$-local and $J$-extensive operator. 
For an arbitrary  $k_A$-local operator $A$, the norm of the commutator $[H,A]$ is bounded from above by
 \begin{align}
\|[H,A]\| \leq 6J k  k_A \| A\|.
\label{eq:bar_inequality}
 \end{align} 
\end{lemma}

The third lemma gives an upper bound of $J_A$ for arbitrary multi-commutators: 
\begin{lemma} \label{extensive_multi_commutator2}
Let $\{A_i\}_{i=1}^n$ be $k_i$-local with $g_i$ extensiveness, respectively. 
Then, the extensiveness of the multi-commutator 
$\{A_n,A_{n-1},\ldots , A_1\}$, which we denote by $J_{\{A_n,A_{n-1},\ldots , A_1\}}$, is bounded from above by
\begin{align}
&J_{\{A_n,A_{n-1},\ldots , A_1\}} \le J_1  \prod_{m=2}^{n} (2  J_{m} \tilde{K}_m)   \label{n_th_commutator_extensive0}
\end{align}
with $\tilde{K}_m := \sum_{i\le {m}} k_i$.
\end{lemma}

{~}\\
\textit{Proof of Lemma~\ref{extensive_multi_commutator2}}.
We prove the lemma by the induction.
For $n=2$, we have 
$
[A_2,A_1] = \sum_{X} \sum_{Y: Y\cap X \neq \emptyset}[a_X^{(2)},a_Y^{(1)}] 
= \sum_{Y} \sum_{X: X\cap Y \neq \emptyset}[a_X^{(2)},a_Y^{(1)}] ,
$
and hence obtain 
\begin{align}
J_{[A_2,A_1]} 
 \le& \sum_{X\ni i} \sum_{Y: Y\cap X \neq \emptyset}2 \|a_X^{(2)}\|\cdot \| a_Y^{(1)}\| \notag \\
 &+ \sum_{Y\ni i} \sum_{X: X\cap Y \neq \emptyset}2 \|a_X^{(2)}\|\cdot \| a_Y^{(1)}\| \notag \\
\le& 2 (k_1+k_2) J_1 J_2.  \label{single_com_ext}
\end{align}

We then assume that \eqref{n_th_commutator_extensive0} is true for $n\le n_0-1$.
For $n=n_0$, from \eqref{single_com_ext} and $\{A_{n_0},A_{n_0-1}, \ldots , A_1\}=\{A_{n_0},\{A_{n_0-1}, \ldots , A_1\}\} $, we have 
\begin{align}
J_{\{A_{n_0},\ldots , A_1\}}  \le& 2J_{n_0} J_{\{A_{n_0-1},\ldots , A_1\}}\Bigl(k_{n_0}+\sum_{m=1}^{n_0-1}k_m \Bigr)  \notag \\
=& J_1  \prod_{m=2}^{n_0} (2  J_{m} \tilde{K}_m) ,\notag
\end{align}
where  we use the fact that $\{A_{n_0-1}, \ldots A_2, A_1\}$ is at most $(\sum_{m=1}^{n_0-1}k_m)$-local. 
This completes the proof. $\blacksquare$

As a standard example, we apply the above lemmas to the case where $|L|=k$, $k_i=k$ and $J_i=J$ for $i=1,2,\ldots,n$.
Then, Lemma~\ref{norm_multi_commutator1} gives us 
\begin{align}
&\| \{A_n,A_{n-1},\ldots , A_1,O_L \}\| \le \lambda^n n! \|O_L\|, \label{fundamental_ineq_special}
\end{align}
where we use the definition of $\lambda=2kJ$.
Moreover,from Lemma~\ref{extensive_multi_commutator2}, we have
\begin{align}
J_{\{A_n,A_{n-1},\ldots , A_1\}}  \le \lambda^{n-1} n! J. \label{n_th_commutator_extensive}
\end{align}

Finally, we denote that the FM expansion terms $\{\Omega_n\}$ are $k_{\Omega}^{(n)}$-local and $J_{\Omega}^{(n)}$-extensive, respectively; 
from the inequality~\eqref{n_th_commutator_extensive} and Eq.~\eqref{eq:Omega_n}, we have $k_{\Omega}^{(n)}\le (n+1)k$ and
\begin{align}
J_{\Omega}^{(n)} \leq\frac{\lambda^{n} (n+1)! J}{(n+1)^2} = \frac{\lambda^{n} J}{n+1} n!  .
\label{Extensiveness_FM_terms}
\end{align}

\section{Proof of Lemma~\ref{FL:FM_bound}}  \label{FL:proof_FM_bound}

We here show the proof of Lemma~\ref{FL:FM_bound}, which gives the upper bound of $\{\| \Omega_n\|\}_{n=0}^\infty$ in the FM-expansion~\eqref{eq:FM}. 

{~}\\
\textit{Proof of Lemma~\ref{FL:FM_bound}.}
For the proof, we start from the explicit form of the FM terms in Eq.~\eqref{eq:Omega_n}. 
From the inequality~\eqref{fundamental_ineq_special}, We first obtain 
\begin{align}
&\|\{ H(t_{\sigma(n+1)}),\ldots,H(t_{\sigma(2)}),H(t_{\sigma(1)})\}\|   \notag \\
\le& \{ H(t_{\sigma(n+1)}),\ldots,H(t_{\sigma(2)}),V(t_{\sigma(1)})\}\| \notag \\
&+\|\{ H(t_{\sigma(n+1)}),\ldots,H_0,V(t_{\sigma(2)})\} \| \notag \\
\le&  \lambda^n n! \sum_{|X|\le k}   (\|v_X(t_{\sigma(1)})\|+\|v_X(t_{\sigma(2)})\|), \notag 
\end{align}
which yields the upper bound of the norm $\|\Omega_n\|$ as
\begin{align}
\|\Omega_n\| \le&  \frac{\lambda^n n! }{(n+1)^2 T^{n+1}}\int_0^Tdt_{n+1}\dots\int_0^{t_2}dt_1    \notag \\
& \sum_{\sigma} \sum_{|X|\le k}  (\|v_X(t_{\sigma(1)})\|+\|v_X(t_{\sigma(2)})\|) \notag \\
\le & \frac{2 \lambda^n n! }{(n+1)^2 T}  \sum_{|X|\le k}  \int_0^{T}dt  \|v_X(t)\|=\frac{2V_0 \lambda^{n}n!}{(n+1)^2}  ,\notag
\end{align}
where in the equality we use the definition of  \eqref{Notations_V_0_lambda}. 
This completes the proof of Lemma~\ref{FL:FM_bound}. $\blacksquare$

\section{Derivation of the inequalities~(\ref{proof_basic_Phi_n}) and (\ref{bound_for_T_n_Phi_n_0})} \label{Sec:tilde_Gamma_l_S}
We here evaluate the norms $\| \Phi_{1,n}\|$ and $\| \Phi_{1,n}^{(n_0)}\|$ which are defied in  Eq.~\eqref{expansion_U_1_U_1_n0}. 
It is enough to prove the inequality for $U_1$ and $U^{(n_0)}_1$; we can prove the other cases in the same way.

We first calculate $\| \Phi_{1,n}\|$. From Eq.~\eqref{unitary_decomp_total_detail}, it can be given by the expansion of
\begin{align}
U_1&= \mathcal{T} \bigl[e^{-i\int_0^T U^\dagger_{\flo,1}(t) V_1(t) U_{\flo,1}(t) dt} \bigr] \notag \\
&=\mathcal{T} \bigl[e^{ -i\int_0^T  \sum_{m=0}^\infty t^m \Gamma_m(t)  \label{express_U_1_T_with_respect_to}
 dt} \bigr]  
\end{align}
with respect to $T$, where we introduce the notation
\begin{align}
 U_{\flo,1}^\dagger(t) V_1(t) U_{\flo,1}(t) = \sum_{m=0}^\infty t^m \Gamma_m(t) .  \notag 
\end{align}
Each of the terms $\{\Gamma_m(t)\}$ is given by
\begin{align}
\Gamma_m(t) := &\frac{i^m}{t^m} \int_0^t \int_0^{t_1} \cdots   \int_0^{t_{m-1}}dt_1 dt_2 \cdots dt_m\notag \\
&\{\tilde{H}_1(t_m),\tilde{H}_1(t_{m-1}), \ldots \tilde{H}_1(t_1), V_1 (t) \}  \notag , 
\end{align}
where $\{\tilde{H}_1(t_m),\tilde{H}_1(t_{m-1}), \ldots \tilde{H}_1(t_1), V_1 (t) \}$ denotes the multi-commutator.
By the use of \eqref{fundamental_ineq_special} with the definition~\eqref{Definition_of_bar_H_m_t}, we can obtain the inequality of
\begin{align}
\|\Gamma_{m}(t)\| \le \lambda^m\sum_{X:X \ni i, X\cap \Lambda_{i-1}=\emptyset}\| v_X(t)\| =: \lambda^m \Vn_1(t).\label{Def_Vn_1_t}
\end{align}
Note that $\Vn_1(t)\le J$ from \eqref{eq:extensiveness} and $\int_0^T \Vn_1(t)dt =T\Vn_1$ from the definition~\eqref{Def_Vni}.
Using the expression of \eqref{express_U_1_T_with_respect_to},  we have
\begin{widetext}
\begin{align}
 T^n \| \Phi_{1,n} (T)\|&\le \sum_{q\ge 1} \sum_{m_1+\cdots+m_q=n-q} \int_0^T \int_0^{t_1} \cdots \int_0^{t_{q-1}}
 t_1^{m_1}\|\Gamma_{m_1} (t_1)\|  t_2^{m_2}\|\Gamma_{m_2}(t_2) \| \cdots   t_q^{m_q}\|\Gamma_{m_q}(t_q) \| dt_1 dt_2 \cdots dt_q\notag \\
 &\le \sum_{q\ge 1} \sum_{m_1+\cdots+m_q=n-q}  (\lambda T)^{n-q}\int_0^T \int_0^{t_1} \cdots \int_0^{t_{q-1}}
\Vn_1 (t_1) \Vn_1(t_2)   \cdots  \Vn_1 (t_q)  dt_1 dt_2 \cdots dt_q  \notag \\
& \le \sum_{q\ge 1}  (\lambda T)^n  2^{n-1} \frac{(\Vn_1/\lambda)^q}{q!}  \le  \Vn_1T (2\lambda T)^{n-1}  e^{\Vn_1/ \lambda} \le \Vn_1 T  e^{1/ (2k)}  2^{-n+1}    , \label{proof_basic_Phi_n_app}
\end{align}
\end{widetext}
where we use 
$\sum_{m_1+\cdots+m_q=n-q} =\binom{n-1}{n-q} \le 2^{n-1}$ in the third inequality, 
$e^x-1 \le xe^x$ in the fourth inequality, and the condition $T\le 1/(4\lambda)$ in the last inequality.
We thus obtain the inequality~\eqref{proof_basic_Phi_n}.

Second, we calculate $ \| \tilde{\Phi}_{1,n} (T)\|$, which is given by the expansion of 
$
U_1^{(n_0)}= \mathcal{T} \bigl[ e^{-i\int_0^T  e^{i\tilde{H}_1^{(n_0)}t}   V_1^{(n_0)}e^{-i\tilde{H}_1^{(n_0)}t} dt} \bigr]
$
with respect to $T$. 
For $n_0=0$, we trivially obtain $\sum_{n=n_0+1}^{\infty} T^n\| \Phi^{(n_0)}_{1,n}\| =0$ because of $H_\flo^{(n_0)}=H_0$ and $U^{(n_0)}_i=\hat{1}$ for $i=1,2,\ldots,N$; in the following, we consider $n_0\ge1$.
Remembering that 
$V_1^{(n_0)}=\tilde{H}^{(n_0)}_{0}-\tilde{H}^{(n_0)}_1$ is given by the use of the FM expansion for $H(t)$ and $\tilde{H}_1(t)$,
we first expand as
\begin{align}
e^{-i\tilde{H}_1^{(n_0)}t}   V_1^{(n_0)} e^{i\tilde{H}_1^{(n_0)}t}    &=\sum_{l,s=0}^\infty t^s T^l \tilde{\Gamma}_{l,s} . \label{tilde_Gamma_l_S_form}
\end{align}
Then, we obtain a similar inequality to \eqref{proof_basic_Phi_n_app}:
\begin{align}
T^n \| \Phi^{(n_0)}_{1,n} (T)\|  \le& \sum_{q\ge 1} \sum_{m_1+m_2+\cdots+m_q=n-q}  \frac{T^n}{q!} \notag \\
&\times \|\tilde{\Gamma}_{m_1}' \| \cdot \|\tilde{\Gamma}_{m_2}' \| \cdots \|\tilde{\Gamma}_{m_q}' \| ,\label{T_n_phi_n0_expand}
\end{align}
where we define 
$
\|\tilde{\Gamma}_{m}' \|:=\sum_{l+s=m } \|\tilde{\Gamma}_{l,s}\|.
$

In Appendix~\ref{Sec:tilde_Gamma_l_S}, we can prove the following inequality for $ \tilde{\Gamma}_{l,s}$:
\begin{align}
\| \tilde{\Gamma}_{l,s} \| \le  (2\lambda)^{l+s}n_0^l   \Vn_1\binom{l+s}{l}  . \label{tilde_Gamma_l_S}
\end{align}
We thus obtain 
$
\|\tilde{\Gamma}_m'  \| 
\le  \Vn_1\bigl[ 2\lambda (n_0+1)\bigr]^{m} \le  \Vn_1(4\lambda n_0)^{m} ,
$
 where we use $n_0\ge1$.
The inequality~\eqref{T_n_phi_n0_expand} then  reduces to
\begin{align}
& T^n \| \Phi^{(n_0)}_{1,n} (T)\|\le  T^n\sum_{q\ge 1} \sum_{m_1+\cdots+m_q=n-q} (4\lambda n_0)^{n-q}  \frac{\Vn_1^q}{q!} \notag \\
 & \le   \frac{ (8\lambda n_0T)^n}{2}  \sum_{q\ge 1}    \frac{[\Vn_1/(4\lambda n_0)]^q}{q!} 
 \le  \Vn_1 T e^{1/(8k) } 2^{-n+1} ,\notag 
\end{align}
where we use $\sum_{m_1+\cdots+m_q=n-q}  \le 2^{n-1}$ in the second inequality and the definition of $n_0=\lfloor 1/(16\lambda T) \rfloor$ in the last inequality.
This completes the proof of the inequality~\eqref{bound_for_T_n_Phi_n_0}.

\subsection{Proof of the inequality~(\ref{tilde_Gamma_l_S})} \label{Sec:tilde_Gamma_l_S}
Remembering that $\tilde{H}_0^{(n_0)}$ and $\tilde{H}_1^{(n_0)}$ come from the FM expansion with respect to $H(t)$ and $\tilde{H}_1(t)$, respectively as in~\eqref{Definition_of_bar_H_m_t_n0}, we first expand $V_1^{(n_0)}=\tilde{H}_0^{(n_0)}-\tilde{H}_1^{(n_0)}$ as 
\begin{align}
 V_1^{(n_0)}= \sum_{q=0}^{n_0}T^q R_q = \sum_{q=0}^{n_0}T^q (\Omega_q-\Omega'_q ) ,
\end{align}
where we denote the FM terms with resect to the Hamiltonian $\tilde{H}_1(t)$ by  $\{\Omega'_{n}\}$, namely $\tilde{H}_1^{(n_0)} = \sum_{n=0}^{n_0}T^n \Omega'_{n}$.
Using Eq.~\eqref{eq:Omega_n}, the terms $R_q$ is explicitly given by  
\begin{widetext}
 \begin{align}
&R_q=\frac{1}{(q+1)^2}\sum_{\substack{i_1,i_2,\ldots,i_{q+1}=0,1 \\ \{i_1,i_2,\ldots, i_{q+1}\} \neq \{1,1,\ldots, 1\} } } \sum_{\sigma}(-1)^{q-\theta(\sigma)} \frac{\theta(\sigma)!(q-\theta(\sigma))!}{q!}
\nonumber \\
&\times\frac{1}{i^q T^{q+1}}\int_0^Tdt_{q+1}\int_0^{t_{q+1}}dt_n\dots\int_0^{t_2}dt_1
\{M_{i_{q+1}}(t_{\sigma(q+1)}),M_{i_q}(t_{\sigma(q)}),\dots,M_{i_2}(t_{\sigma(2)}),M_{i_1}(t_{\sigma(1)})\} ,
\label{Explicit_R_q}
\end{align}
where $\{M_{i_{q+1}}(t_{\sigma(q+1)}),M_{i_q}(t_{\sigma(q)}),\dots,M_{i_2}(t_{\sigma(2)}),M_{i_1}(t_{\sigma(1)})\}$ denotes the multi-commutator, $M_0(t):=V_1(t)$ and $M_1(t):= \tilde{H}_1(t)$; note that $H(t)=V_1(t)+\tilde{H}_1(t)$ from the definition~\eqref{Definition_of_bar_H_m_t}. 
  
We, in the following, calculate the expansion of $ e^{-i\tilde{H}_1^{(n_0)}t}   V_1^{(n_0)} e^{i\tilde{H}_1^{(n_0)}t}$:
\begin{align}
e^{-i\tilde{H}_1^{(n_0)}t}   V_1^{(n_0)} e^{i\tilde{H}_1^{(n_0)}t}=  \sum_{s=0}^\infty  \sum_{l=0}^\infty T^l \sum_{q\le l} \frac{(-it)^s}{s!} \sum_{q_1+q_2+\cdots +q_s=l-q} \{ \Omega'_{q_s}, \Omega'_{q_{s-1}}, \ldots , \Omega'_{q_1},R_q \}.
 \label{expansion_of_tilde_gamma_l_s_form}
\end{align}
\end{widetext}
From the expansion~\eqref{expansion_of_tilde_gamma_l_s_form}, we give the explicit form of $ \tilde{\Gamma}_{l,s}$ in Eq.~\eqref{tilde_Gamma_l_S_form} as
\begin{align}
\tilde{\Gamma}_{l,s}=\sum_{q\le l}& \sum_{q_1+q_2+\cdots +q_s=l-q} \frac{i^s}{s!}\{ \Omega'_{q_s}, \Omega'_{q_{s-1}}, \ldots , \Omega'_{q_1},R_q \}. \label{Form_tilde_Gamma_l_s_} 
\end{align}

In order to  evaluate the norm of $\|\{ \Omega'_{q_s}, \Omega'_{q_{s-1}}, \ldots , \Omega'_{q_1},R_q \}\|$, we first derive the following inequality:
\begin{align}
&\| \{ \Omega'_{q_1}, M_{i_{q+1}}(t_{q+1}),M_{i_q}(t_{q}),\dots,M_{i_1}(t_{1})\}  \| \notag \\
 & \le \frac{( q+1) \lambda^{q_1+q+1}  q_1! q!}{q_1+1}\Vn_1(t_{s+1}) \label{multicommutator_norm_1}
\end{align}
for $\{i_1,i_2,\ldots, i_{q+1}\} \neq \{1,1,\ldots, 1\}$, 
where $\Vn_1(t)$ is defined in Eq.~\eqref{Def_Vn_1_t} and we denote $s$ such that $i_{s+1}=0$. 
For the proof, we define 
$
\tilde{M}_s:= \{M_{i_{s}}(t_{s}) , \ldots , M_{i_1}(t_{1})\}  ,
$
and obtain 
\begin{align}
& \{ \Omega'_{q_1}, M_{i_{q+1}}(t_{q+1}),M_{i_q}(t_{q}),\ldots, M_{i_1}(t_{1})\}  \notag \\
 &=\{ \Omega'_{q_1}, M_{i_{q+1}}(t_{q+1}),\ldots , M_{i_{s+1}}(t_{s+1}) ,\tilde{M}_s  \}  \notag \\
&= -\{ \Omega'_{q_1}, M_{i_{q+1}}(t_{q+1}),\ldots \tilde{M}_s , V_1(t_{s+1}) \} , \label{inequality_B6_appe}
\end{align}
where we use $i_{s+1}=0$ or $M_{i_{s+1}}(t_{s+1}) = V_1(t_{s+1})$ in the second equality.
Because of the inequalities~\eqref{n_th_commutator_extensive} and \eqref{Extensiveness_FM_terms}, we have
\begin{align} 
&\Omega'_{q_1}: \textrm{$[(q_1+1)k]$-local}, \textrm{$J_{\Omega}^{(q_1)}$-extensive}, \notag \\
&M_{i}(t_i):\textrm{$k$-local},\ \textrm{$J$-extensive}, \notag \\
&\tilde{M}_s : \textrm{$(sk)$-local},\ \textrm{$[J\lambda^{s-1}s!]$-extensive} , \notag 
\end{align}
and hence we obtain the inequality~\eqref{multicommutator_norm_1} by applying the inequality~\eqref{fundamental_ineq} to \eqref{inequality_B6_appe} with 
$J_{\Omega}^{(n)}=\lambda^{n} J n!/(n+1) $.

From \eqref{Explicit_R_q} and \eqref{multicommutator_norm_1}, we have
\begin{align}
\| \{ \Omega'_{q_1}, R_q \}  \| \le \Vn_1 \frac{2^{q+1} (q+1) \lambda^{q_1+q+1}  q_1! q!}{(q_1+1)(q+1)^2} ,
\end{align}
where we use $\int_0^T \Vn_1(t)/T dt =\Vn_1$ and the coefficient $2^{q+1}$ comes from the summation with respect to $\{i_1,i_2,\ldots, i_{q+1}\} \neq \{1,1,\ldots, 1\}$ in Eq.~\eqref{Explicit_R_q}.
In the same way, we can derive the inequality
\begin{align}
&\| \{\Omega'_{q_s}, \Omega'_{q_{s-1}}, \ldots, \Omega'_{q_1}, R_q \} \|  \notag \\
\le & \Vn_1\frac{2^{q+1} \lambda^{q}  q!}{(q+1)^2} \frac{(q+1) \lambda^{q_1+1} q_1!}{q_1+1}\frac{(q+q_1+2) \lambda^{q_2+1} q_2!}{q_2+1}\notag \\
& \cdots  \frac{(q+q_1+q_2+\cdots + q_{s-1} +s) \lambda^{q_s+1} q_s!}{q_s+1} \notag \\
\le &\Vn_1  2^{q+1} \lambda^{q+s}  q! \lambda^{q_1+q_2+\cdots +q_s}q_s!q_{s-1}!\cdots q_1!   \notag \\
&  (q+q_1+2)\cdots (q+q_1 + q_{2} + \cdots +q_{s-1}+s)\notag \\
\le &\Vn_1 2^{q+1}\lambda^{l+s}n_0^l  (l+2) \cdots (l+s), \label{Omega_q_s_q_s1}
\end{align}
where in the last inequality we use the conditions $q\le n_0$, $q_j\le n_0$ ($j=1,2,\ldots,s$) and $q_1+q_2+\cdots +q_s=l-q$, which yield the inequalities $q! q_s!q_{s-1}!\cdots q_1! \le n_0^{q+q_s+q_{s-1} + \cdots + q_1} = n_0^l $ and $q+q_1+q_{2}+ \cdots +q_{j-1}+ j \le l+j$ for $j=1,2,\ldots, s$.

From Eq.~\eqref{Form_tilde_Gamma_l_s_} and the inequality~\eqref{Omega_q_s_q_s1}, we finally obtain the inequality~(\ref{tilde_Gamma_l_S}) as 
\begin{align}
\|\tilde{\Gamma}_{l,s}\| \le& \frac{\lambda^{l+s} \Vn_1}{s!} n_0^l \sum_{q\le l} 2^{q+1}  2^{l+s-q-1}   (l+2) \cdots (l+s)\notag \\
=& \frac{\lambda^{l+s} \Vn_1}{s!} n_0^l  2^{l+s} (l+1)(l+2) \cdots (l+s) \notag \\
=&  (2\lambda)^{l+s}n_0^l   \Vn_1 \binom{l+s}{l} ,\notag 
\end{align}
where we use $\sum_{q_1+q_2+\cdots +q_s=l-q}=\binom{ l+s -q-1}{ l-q } \le 2^{l+s-q-1}$ in the first inequality and $\sum_{q\le l}=l+1$ in the second equality, respectively.


\section{Proof of Theorem~\ref{FL:Heat absorption}}  \label{FL:proof_energy_localization}

The theorem gives the upper bound of the norm $\| \Pi_{\ge E+\Delta E}^{(n)} U_\flo^m  \Pi_{\le E}^{(n)} \|^2$.
Then, we start from the inequality
\begin{align}
&\| \Pi_{\ge E+\Delta E}^{(n)} U_\flo^m  \Pi_{\le E}^{(n)} \|  \notag \\
&= \| \Pi_{\ge E+\Delta E}^{(n)}e^{-\tau H_\flo^{(n)}} e^{\tau H_\flo^{(n)}}  U_\flo^m e^{-\tau H_\flo^{(n)}}  e^{\tau H_\flo^{(n)}} \Pi_{\le E}^{(n)} \|  \notag \\
&\le e^{-\tau \Delta E} \| e^{\tau H_\flo^{(n_0)} } U_\flo e^{-\tau H_\flo^{(n_0)} } \|^m  \notag \\ 
&\quad  \times \|e^{\tau H_\flo^{(n)}} e^{-\tau H_\flo^{(n_0)} } \| \cdot\| e^{\tau H_\flo^{(n_0)} } e^{-\tau H_\flo^{(n)}} \|  ,
 \label{DeltaU_energy_absorption_localize1}
\end{align}
where we use $\| \Pi_{\ge E+\Delta E}^{(n)}e^{-\tau H_\flo^{(n)}}\| \le e^{-\tau (E+\Delta E)}$ and $\|e^{\tau H_\flo^{(n)}} \Pi_{\le E}^{(n)} \|\le e^{\tau E}$.
We therefore evaluate the three norms: $\|e^{\tau H_\flo^{(n)}} e^{-\tau H_\flo^{(n_0)}}\|$, $\|e^{\tau H_\flo^{(n_0)}} e^{-\tau H_\flo^{(n)}}\|$ and $\|e^{\tau H_\flo^{(n_0)}} U_\flo e^{-\tau H_\flo^{(n_0)}}\|$.
We first show the sketch and give the technical details in Appendixes~\ref{Der:first_ineq_theorem3} and \ref{Der:second_ineq_theorem3}.

First, to evaluate the norm of $\|e^{\tau H_\flo^{(n_0)} } e^{-\tau H_\flo^{(n)}} \| $, we use 
\begin{align}
e^{-\tau H_\flo^{(n)}}  = e^{-\tau H_\flo^{(n_0)} }\mathcal{T}\Bigl[  e^{\int_0^\tau e^{xH_\flo^{(n_0)}}( H_\flo^{(n_0)}-H_\flo^{(n)})  e^{-xH_\flo^{(n_0)}} dx }\Bigr], \notag 
\end{align}
which yields
\begin{align}
&\|e^{\tau H_\flo^{(n_0)} } e^{-\tau H_\flo^{(n)}} \|  \notag \\
\le& \exp \biggl( \int_0^\tau \sum_{m=n+1}^{n_0}T^m \| e^{xH_\flo^{(n_0)}}\Omega_m  e^{-xH_\flo^{(n_0)}}\|   dx \biggr) .  \label{DeltaU_energy_localize1_0}
\end{align}
By using Lemma~\ref{FL:FM_bound} on the norm of $\Omega_m$, we obtain the upper bound of 
\begin{align}
&\|e^{\tau H_\flo^{(n)}} e^{-\tau H_\flo^{(n_0)}} \|  \le e^{\tau V_0 T^{n+1}\tilde{W}_{n+1}/2} ,\label{first_ineq_theorem3}
\end{align}  
with the derivation in Appendix~\ref{Der:first_ineq_theorem3}, where $\tilde{W}_{n+1}$ is defined in the theorem: $\tilde{W}_{n}=2 (4\lambda/3)^n  n! /(n+1)^2$.
For $\|e^{\tau H_\flo^{(n_0)}} e^{-\tau H_\flo^{(n)}}\|$, we can also obtain the same inequality.

We then consider the norm of 
\begin{align}
\|e^{\tau H_\flo^{(n_0)}} U_\flo e^{-\tau H_\flo^{(n_0)}}\|\le \|e^{\tau H_\flo^{(n_0)}} U_\flo U^{(n_0)\dagger}_\flo e^{-\tau H_\flo^{(n_0)}}\|, \notag 
\end{align}  
where we use $[U^{(n_0)}_\flo, H_\flo^{(n_0)}]=\hat{0}$, $\|U^{(n_0)}_\flo \|=1$ and the inequality $\|A B\| \le \|A\| \cdot \|B\|$ for arbitrary operators $A$ and $B$.
We here adopt the decompositions \eqref{unitary_decomp_total} and \eqref{unitary_decomp_total_effective}, and introduce the notations of
$
s_i= \|e^{\tau H_\flo^{(n_0)}} U_{\flo,i}(T) U_{\flo,i}^{(n_0)\dagger}(T) e^{-\tau H_\flo^{(n_0)}}\|
$
for $i=0,1,\ldots,N$.  
Because of $U_{\flo,i-1}(T)=U_{\flo,i}(T)U_{i}$ from Eq.~\eqref{unitary_decomp_total_detail}, we have 
\begin{align}
s_{i-1}& \le \|e^{\tau H_\flo^{(n_0)}} U_{\flo,i}(T) U_{\flo,i}^{(n_0)\dagger}(T) e^{-\tau H_\flo^{(n_0)}}\| \notag \\
+& \|e^{\tau H_\flo^{(n_0)}} U_{\flo,i}(T) (\hat{1}- U_{i}  U^{(n_0)\dagger}_{i} ) U^{(n_0)\dagger}_{\flo,i}(T)  e^{-\tau H_\flo^{(n_0)}}\| \notag   \\
&\le s_{i} + 44\Vn_i J T^2  2^{-n_0/2},\label{0second_ineq_theorem3}
\end{align}  
where the second inequality comes from lemma~\ref{FL:Metastability_local} which ensures $U_{i}  U^{(n_0)\dagger}_{i}\simeq \hat{1}$; we give the derivation in Appendix~\ref{Der:second_ineq_theorem3}.
From the definition of $s_i$, we have $s_0=\|e^{\tau H_\flo^{(n_0)}} U_\flo U^{(n_0)\dagger}_\flo  e^{-\tau H_\flo^{(n_0)}}\|$ and $s_N=1$. Hence, from Eq.~\eqref{summation_Vn_i},
the inequality~\eqref{0second_ineq_theorem3} reduces to, 
\begin{align}
&\|e^{\tau H_\flo^{(n_0)}} U_\flo e^{-\tau H_\flo^{(n_0)}}\|\le s_0\le 1+ 22\tau V_0 \lambda  T  2^{-n_0/2},  \label{second_ineq_theorem3}
\end{align}  
where we use the assumption $T\le \tau$ and $J=\lambda/(2k)\le \lambda/2$.

By applying the inequalities~\eqref{first_ineq_theorem3} and \eqref{second_ineq_theorem3} to the inequality~\eqref{DeltaU_energy_absorption_localize1}, 
we prove Theorem~\ref{FL:Heat absorption}. $\blacksquare$

\subsection{Derivation of the inequality \eqref{first_ineq_theorem3}} \label{Der:first_ineq_theorem3}

We start from the inequality~\eqref{DeltaU_energy_localize1_0} and evaluate the norm of $\| e^{xH_\flo^{(n)}}\Omega_m  e^{-xH_\flo^{(n)}}\|$.
For the purpose, we utilize the following technical lemma:

\begin{lemma}\label{FL:evaluate_floquet_ham}
Let $A$ be an arbitrary $(n_Ak)$-local operator such that 
$
A=\sum_{|X|\le n_Ak} a_X 
$.
Then, the norm of $\|e^{xH_\flo^{(n_0)}}A  e^{-xH_\flo^{(n_0)}}\|$ is bounded from above by:
\begin{align}
\|e^{xH_\flo^{(n_0)}} A e^{-x H_\flo^{(n_0)}}\|\le  \frac{\eta^{n_A}}{2( 1-2 \eta \lambda n_0 T) }\|A\|, \label{Inequality_for_A_Norm}
\end{align} 
where $\eta :=1/( 1-2\tilde{\lambda}x)$ and $\tilde{\lambda}:=6k^2J$. 
The proof is given below. 
\end{lemma}

From Eq.~\eqref{eq:Omega_n}, the operator $\Omega_m$ is given by the sum of $(m+1)$th multi-commutators, and hence 
it contains at most $(m+1)k$-body couplings, namely $n_A=m+1$. 
For $x\le \tau=1/(8\tilde{\lambda})$ and $n_0 = \lfloor1/(16\lambda T) \rfloor$, we have
$\eta \le 4/3$ and $1/[2( 1-2 \eta \lambda n_0 T)]\le 3/5$. 
Therefore, using Lemma~\ref{FL:evaluate_floquet_ham} with Lemma~\ref{FL:FM_bound}, we obtain
\begin{align}
&\sum_{m=n+1}^{n_0}T^m \| e^{xH_\flo^{(n_0)}}\Omega_m  e^{-xH_\flo^{(n_0)}}\| \le \frac{4}{5} \sum_{m=n+1}^{n_0}\Bigl(\frac{4T}{3}\Bigr )^{m}\overline{\Omega}_m  \notag \\
&\le \frac{V_0  (4\lambda T/3)^{n+1}}{(n+2)^2} (n+1)!  = \frac{V_0 T^{n+1}\tilde{W}_{n+1}}{2}  \label{1DeltaU_energy_localize1_02}
\end{align}
in the similar way to the derivation of \eqref{eq:general_n_proof2}. 

By combining the inequalities \eqref{DeltaU_energy_localize1_0} and \eqref{1DeltaU_energy_localize1_02}, we obtain 
$\|e^{\tau H_\flo^{(n_0)} } e^{-\tau H_\flo^{(n)}} \| \le e^{\tau V_0 T^{n+1}\tilde{W}_{n+1} /2 }$. The same inequality can be given for $\|e^{\tau H_\flo^{(n)}} e^{-\tau H_\flo^{(n_0)} }\|$. 
We thus prove the inequality in \eqref{first_ineq_theorem3}.

{~}\\
\textit{Proof of Lemma~\ref{FL:evaluate_floquet_ham}}. 
We here calculate an upper bound of
$
\|e^{x H_\flo^{(n_0)}} A e^{-x H_\flo^{(n_0)}}\| ,
$
where $A$ is an arbitrary $(n_A k )$-local operator due to the condition in Lemma~\ref{FL:evaluate_floquet_ham}.
We have
\begin{align}
\|e^{x H_\flo^{(n_0)}} A e^{-x H_\flo^{(n_0)}}\|\le  &\sum_{s=0}^\infty \frac{x^s}{s!}\sum_{M=0}^{s n_0} T^M  \sum_{\substack{m_1+m_2+\ldots+m_s=M \\\{ m_j\le n_0\} }} \notag \\ 
&\| \{\Omega_{m_s},\Omega_{m_{s-1}},\ldots, \Omega_{m_1},A\}\|.\notag
\end{align} 
Under the assumption of $m_1+m_2+\cdots+m_s=M$, we use the inequality~\eqref{eq:bar_inequality} and obtain 
\begin{widetext}
\begin{align}
&\| \{\Omega_{m_s},\Omega_{m_{s-1}},\ldots, \Omega_{m_1},A\}\|\notag \\
 \le&  \prod_{j=1}^s[6J_{\Omega}^{(m_j)}  k^2 (m_j+1)] n_A  (n_A+ m_1 +1) (n_A+ m_1 +m_2+2) \cdots  (n_A+ m_1 +m_2+\cdots +m_{s-1} +s-1) \|A\|\notag \\
 \le &  (6 J k^2)^s \lambda^{m_1+m_2+\cdots+m_s} m_1! m_2! \cdots m_s!   n_A (n_A+ m_1 +1) \cdots  (n_A+ m_1 +m_2+\cdots +m_{s-1} +s-1)\|A\|  \notag \\
 \le &\tilde{\lambda}^s (\lambda n_0)^M (n_A+ M ) (n_A+M +1) \cdots  (n_A+ M+s-1)\|A\| ,\label{The_inequatliy_D1_egf}
\end{align} 
where we use the fact that $\Omega_m$ is $[(m+1)k]$-local and $J_{\Omega}^{(m)}$-extensive with $J_{\Omega}^{(m)} \le \lambda^m J m! /(m+1)$ as in \eqref{n_th_commutator_extensive}, and $m! \le n_0^m$ for $m\le n_0$.
We thus obtain the inequality~\eqref{Inequality_for_A_Norm} as follows:
\begin{align}
\|e^{x H_\flo^{(n_0)}} A e^{-x H_\flo^{(n_0)}}\|
&\le  \|A\|\sum_{M=0}^{\infty}  \sum_{s=0}^\infty \frac{(2\tilde{\lambda} x)^s}{s!}\frac{ (2\lambda n_0 T)^M}{2} (n_A+ M ) (n_A+M +1) \cdots  (n_A+ M+s-1)\notag \\
&= \|A\| \frac{ (1-2\tilde{\lambda}x)^{-n_A}}{2} \sum_{M=0}^{\infty} \left ( \frac{2\lambda n_0 T}{1-2\tilde{\lambda}x }\right)^M
= \frac{\eta^{n_A}  }{2( 1-2 \eta \lambda n_0 T) }\|A\|, \notag 
\end{align} 
\end{widetext}
where the first inequality comes from \eqref{The_inequatliy_D1_egf} and $\sum_{m_1+m_2+\ldots+m_s=M}\le \binom{M+s-1}{M} \le 2^{M+s-1}$, the first equality comes from the Taylor expansion of $(1-2\tilde{\lambda}x)^{-(n_A+ M)}$ and  we use the definition of $\eta :=1/( 1-2\tilde{\lambda}x)$ in the second equality. 
This completes the proof of the lemma. $\blacksquare$

\subsection{Derivation of the inequality~\eqref{0second_ineq_theorem3}} \label{Der:second_ineq_theorem3}

We here derive the upper bound of 
$
\Lambda_i:= \|e^{\tau H_\flo^{(n_0)}} U_{\flo,i}(T)(  U_{i}  U_{i}^{(n_0)\dagger} -\hat{1}) U_{\flo,i}^{\dagger}(T)  e^{-\tau H_\flo^{(n_0)}}\|
$
as  $\Lambda_i \le   44\Vn_i J T^2  2^{-n_0/2} $.
For this purpose, we first expand $U_{i}  U_{i}^{(n_0)\dagger}$ with respect to $T$ and obtain
\begin{align}
U_{i}  U_{i}^{(n_0)\dagger} = \sum_{m,m'=0}^{\infty} T^{m+m'} \Phi_{i,m} \Phi_{i,m'}^{(n_0)} =   \sum_{n=0}^\infty T^n \Psi_{i,n},\notag 
\end{align} 
where we use the notations in \eqref{expansion_U_1_U_1_n0} and define $\Psi_{i,n}:=  \sum_{m+m'=n} \Phi_{i,m} \Phi_{i,m'}^{(n_0)}$.
Because $ U_{i} U_{i}^{(n_0)\dagger}   $ is equal to $ U_{i}U_{i}^{\dagger}  $ up to the order of $T^{n_0}$, we have
\begin{align}
 U_{i}  U_{i}^{(n_0)\dagger} =\hat{1}+  \sum_{n\ge n_0+1} T^n \Psi_{i,n},
\end{align} 
and hence we obtain
\begin{align}
 \Lambda_i &\le  \sum_{n\ge n_0+1} T^n   \Lambda_{i,n}, \label{heat_ab_R_j0} 
\end{align} 
where $\Lambda_{i,n}:=\|e^{\tau H_\flo^{(n_0)}} U_{\flo,i}(T) \Psi_{i,n}U_{\flo,i}^\dagger(T)  e^{-\tau H_\flo^{(n_0)}}\|$.

We prove the following inequality below:
\begin{align}
\Lambda_{i,n} \le \frac{3}{5}\|\Psi_{i,n}\| 2^{n/2}   \label{heat_ab_R_j1}
\end{align} 
for $\tau = 1/(8\tilde{\lambda})$.
Moreover, by following the derivation of the inequalities~\eqref{proof_basic_Phi_n} and \eqref{bound_for_T_n_Phi_n_0} with $T\le 1/(8\tilde{\lambda}) \le 1/(8\lambda) $, we have
\begin{align}
&T^m \| \Phi_{i,m} (T)\| \le  \Vn_i T e^{1/(2k)} 2^{-2 m+2} ,\notag \\
&T^m\| \Phi_{i,m}^{(n_0)} (T)\| \le \Vn_i T e^{1/(8k)} 2^{- m+1}, \notag 
\end{align} 
which yields
\begin{align} 
&T^n \| \Psi_{i,n}\| \le   \sum_{m+m'=n} \|T^m \Phi_{i,m}\| \cdot \|T^{m'}\Phi_{i,m'}^{(n_0)} \| \notag \\
&\le15 \Vn_i^2T^2  \sum_{m+m'=n} 2^{-2m-m'}\le30\Vn_i J T^2   2^{-n}, \label{T_n_Psi_j_n_heat}
\end{align} 
where we use $8e^{5/8} \le15$ and $\Vn_j \le J$.

By combining the inequalities \eqref{heat_ab_R_j1} and \eqref{T_n_Psi_j_n_heat} with \eqref{heat_ab_R_j0}, we obtain
\begin{align}
 \Lambda_i  \le&\frac{3}{5}  \sum_{n\ge n_0+1}  2^{n/2}T^n \|\Psi_{i,n}\| \le 18 \Vn_i J T^2 \sum_{n\ge n_0+1} 2^{-n/2} , \notag  
\end{align} 
which finally reduces to the inequality of~$\Lambda_i \le   44\Vn_i J T^2  2^{-n_0/2} $.

{~}\\
\textit{Derivation of the inequality~\eqref{heat_ab_R_j1}.}
For the evaluation of $\Lambda_{i,n}$, we first expand as
\begin{align}
U_{\flo,i}(T) \Psi_{i,n} U_{\flo,i}^\dagger(T) = \sum_{q=0}^\infty T^q\tilde{\Psi}_{i,n+q}  \label{Norm_R_j1_Psi_R_j1}
 \end{align} 
and evaluate the norm of $\tilde{\Psi}_{i,n+q}$. From the definition~\eqref{unitary_decomp_total_detail} of $U_{\flo,i}(T)$, we obtain
\begin{align}
T^q \tilde{\Psi}_{i,n+q} =(- i)^q &\int_0^T \int_0^{t_1} \cdots \int_0^{t_{q-1}} dt_1 dt_{2}\cdots dt_q \notag \\
&\{\tilde{H}_{i}(t_1),\tilde{H}_{i}(t_{2}),\ldots, \tilde{H}_{i}(t_q),\Psi_{i,n}\} . \notag 
\end{align} 

Because $\tilde{H}_{i}(t_1)$ is $k$-local and $J$-extensive, the inequality~\eqref{eq:bar_inequality} gives
\begin{align}
&\|  \{\tilde{H}_{i}(t_1),\tilde{H}_{i}(t_{2}),\ldots, \tilde{H}_{i}(t_q),\Psi_{i,n}\} \| \notag \\
&\le \tilde{\lambda}^q  n(n+1)\cdots (n+q-1)\|\Psi_{i,n}\|,\notag 
\end{align} 
where $\tilde{\lambda}=6k^2J$ and use the fact that $\Psi_{i,n}$ is at most $(nk)$-local.
We thus obtain
\begin{align}
T^q \| \tilde{\Psi}_{i,n+q} \| &\le (\tilde{\lambda}T)^q  \frac{n(n+1) \cdots (n+q-1)}{q!}  \|\Psi_{i,n}\| .\label{T_q_tilde_psi_j_nq}
\end{align} 
By substituting $A$ with $\Psi_{i,n+q}$ in the inequality~\eqref{Inequality_for_A_Norm} of Lemma~\ref{FL:evaluate_floquet_ham}, we have
\begin{align}
 \|e^{\tau H_\flo^{(n_0)}} \tilde{\Psi}_{i,n+q}  e^{-\tau H_\flo^{(n_0)}}\|  \le \frac{\eta^{n+q} \| \tilde{\Psi}_{i,n+q}\| }{2( 1-2 \eta \lambda n_0 T) } \label{T_q_tilde_psi_j_nq2}
\end{align} 
with $\eta=1/( 1-2\tilde{\lambda}\tau)$. 
Thus, from the inequalities \eqref{T_q_tilde_psi_j_nq} and \eqref{T_q_tilde_psi_j_nq2},  we obtain
\begin{align}
&\Lambda_i=\|e^{\tau H_\flo^{(n_0)}} U_{\flo,i}(T) \Psi_{i,n}U_{\flo,i}^\dagger(T)  e^{-\tau H_\flo^{(n_0)}}\| \notag \\
&\le  \sum_{q=0}^\infty \frac{(\tilde{\lambda}T)^q  \eta^{n+q} }{2( 1-2 \eta \lambda n_0 T) }  \frac{n(n+1)  \cdots (n+q-1)}{q!}\|\Psi_{i,n}\| \notag \\
&= \frac{\|\Psi_{i,n}\| }{2( 1-2 \eta \lambda n_0 T) }  \left(\frac{\eta}{1-\eta  \tilde{\lambda}T}\right)^n ,\label{Inequality_for_psi_j_n_Norm}
\end{align} 
where we use the fact that $\tilde{\Psi}_{i,n+q}$ is $[k(n+q)]$-local in the second inequality 
and the last equality comes from the Taylor expansion of $(1-\eta  \tilde{\lambda}T)^{-n}$ with respect to $\eta  \tilde{\lambda}T$.
We here set $\tau = 1/(8\tilde{\lambda})$, $n_0 =\lfloor 1/(16\lambda T) \rfloor $ and $T\le1/(8 \tilde{\lambda})$.
Hence, we have
$
\eta = 4/3,
$
$\frac{\eta}{1-\eta  \tilde{\lambda}T} \le \frac{4/3}{1-1/6}=\frac{8}{5} < 2^{1/2}$ and 
$\frac{1}{2( 1-2 \eta \lambda n_0 T) }  \le 3/5$,
from which the inequality~\eqref{Inequality_for_psi_j_n_Norm} reduces to \eqref{heat_ab_R_j1}.


\bibliography{FL_metastability.bib}

\end{document}